\documentclass[amsmath,amssymb,floatfix,prb,epsf,showpacs,twocolumn]{revtex4-1}
\usepackage{amsmath,amssymb,natbib,bm,graphicx,url,epsfig}
\usepackage[ansinew]{inputenc}
\usepackage{bm}
\usepackage{color}

\newcommand{\be}{\begin{equation}}
\newcommand{\ee}{\end{equation}}
\newcommand{\bes}{\begin{equation}\begin{split}}
\newcommand{\ees}{\end{split}\end{equation}}
\newcommand{\bea}{\begin{eqnarray}}
\newcommand{\eea}{\end{eqnarray}}
\newcommand{\nn}{\nonumber}

\def\beq{\begin{equation}}
\def\eeq{\end{equation}}
\def\bea{\begin{eqnarray}}
\def\eea{\end{eqnarray}}

\begin{document}

\title{ Composite fermion state of spin-orbit coupled bosons}

\author{Tigran A. Sedrakyan$^{1}$, Alex Kamenev$^{1}$, Leonid I. Glazman$^{2}$}

\affiliation{$^1$William I. Fine Theoretical Physics Institute and Department of Physics, University of Minnesota, Minneapolis, MN 55455, USA}

\affiliation{$^2$ Department of Physics, Yale University, New Haven, Connecticut 06520, USA}

%\date{\currenttime \today}
\date{\today}
\begin{abstract}
We consider spinor Bose gas with the isotropic Rashba spin-orbit coupling in 2D.
We argue that at low density its groundstate is a composite fermion state with a Chern-Simons gauge field and filling factor one.
The chemical potential of such a state
scales with the density as $\mu \propto n^{3/2}$.  This is a lower energy per particle than $\mu\propto n$ for the earlier suggested groundstate candidates: a condensate with broken time-reversal symmetry and a spin density wave state.
\end{abstract}

\maketitle

\section{Introduction}

 Precise control over interactions of ultracold atoms with laser fields opened an opportunity of  fabricating
synthetic non-Abelian gauge fields\cite{Lin-2008, Lin-2009, synthetic} and
 spin-orbit (SO) couplings\cite{Lin-2011, zhang-exp}
of both Rashba\cite{Rashba} and Dresselhaus\cite{Dresselhaus} type.
In analogy with previously known vector condensates\cite{stenger, ho, law,ashhab,svist}, an important tool for synthesizing such systems is the controlled Raman coupling between oriented sequence of initially degenerate hyperfine states of $^{87}$Rb.
Interactions of spatially varying laser field with  hyperfine atomic levels effectively produces synthetic non-Abelian gauge fields,\cite{Jaksch, Ruseckas, Osterloh, Zhu}  originating from the Berry phase, and SO coupling\cite{Lin-2011, zhang-exp, tudor, SAG, dalibard, Campbell-2011, dalibard2,3D}  of momentum to the internal isospin degrees of freedom.
Bosons with SO coupling, discussed and studied in the context of cold-atom physics in Ref.~[\onlinecite{SAG}], 
were realized very recently by Spielman's group at NIST\cite{Lin-2011}.
%(we note that these isospin degrees of freedom are not spins in a sense of representation of the $3D$ rotation group, but are rather a product of %"synthetic" $SU(2)$-symmetry remained after projection  on low energy states).
%The latter was possible only because of the efficient control over the parameters
%of the system, particularly detuning parameter $\delta$ and Raby frequences $\Omega$ of laser beams.
Depending on the particular experimental scheme,  i.e. the choice of atomic states and a sequence of optically induced transitions between them, one may in principle realize various SO Hamiltonians for the projected low-energy states.
Low energy properties of such SO bosonic systems have been discussed in Refs.~[\onlinecite{SAG,zhai,Wu,radic,ho11,Lamacraft,zhai2,yongping,DasSarma,ozawa,ozawa2,pitaevskii}].

While fermions with SO coupling were extensively studied over the last decades, see e.g. Refs.~[\onlinecite{kato, hasan, nagaosa, zutic}]
 for reviews,
relatively little is known about SO bosons. Yet, they offer a number of  fundamental problems which, in some respects, are
more challenging than their fermionic counterparts. Indeed, in many instances SO coupling leads to single-particle dispersion relations
which exhibit multiple minima or even degenerate manifold of minimal energy states.
Fermionic many-body systems  form a unique Fermi sea state on top of such dispersion relation.  On the other hand, bosons tend to occupy
the lowest energy states and thus face macroscopic degeneracy of their non-interacting groundstate. It is entirely the effect of
collisions (i.e. boson-boson interactions) which lifts this degeneracy and selects a true many-body groundstate.

In this respect the problem is
somewhat similar to fractional quantum Hall effect (FQHE). There the fermionic kinetic energy is degenerate due to Landau quantization
and the nature of the groundstate is  determined by the interactions. One of the most remarkable concepts, which emerged in FQHE studies,
is that of composite particles\cite{Jain, lopez, Halperin, SDS, heinonen, jbook}. For example, FQHE states with filling fractions $\nu =p/(2p+1)$, where $p=1,2\ldots$, were understood as integer $\nu=p$ quantum Hall states of composite fermion particles. The latter are obtained by binding the original fermions with flux tubes carrying exactly {\em two} flux quanta. Technically such a binding is achieved by assigning Chern-Simons (CS) phase factor to the many-body wavefunction of composite particles.

The goal of this paper is to show that a very similar transformation plays a major role in understanding of the groundstate of 2D bosons with
Rashba SO coupling. We argue here that their groundstate wavefunction may be approximated by that of the Fermi gas at integer filling factor $\nu=1$,
dressed by the Chern-Simons phase with {\em one} flux quantum attached to every fermion. Such phase factor transforms the integer quantum Hall {\em fermionic} wavefunction into a {\em bosonic} one.
Fermionization of the bosonic system allows the latter to minimize its interaction energy. This is due to the fact that fermions with the same spin
can't be at the same spatial point and thus cannot interact through a short-range s-wave interaction. (Consequently the amplitude
for  two fermions with almost parallel spins to be at the same point is small as the angle between their spins.) As the result
the interaction energy per particle in a Fermi sea is smaller than in a Bose-condensate of the same density. For low enough density such
reduction of the interaction energy wins over the associate increase of the kinetic energy.

A very similar physics is behind the so-called Tonks-Girardeau limit of spinless 1D Bose gas\cite{LiebLiniger, tonks, girardeau}. At a small density (which in 1D is the same as strong interactions)  the bosonic wavefunction approaches symmetrized wavefunction of non-interacting Fermi gas. If spin degree of freedom is present in 1D, the groundstate is known to be fully spin-polarized\cite{lieb, yang, gaudin}
again leading to the Tonks-Girardeau construction in low density limit. We thus notice that 2D Rashba bosons share features of both 2D quantum Hall systems as well as 1D Bose gases. The deep connection with the latter is due to the fact that single-particle density of states for particles with Rashba SO coupling behaves as $\epsilon^{-1/2}$ at small energy. This is typical for 1D systems, making particles with Rashba SO coupling ``1D-like'', irrespective of their actual spatial dimensionality.

Technically blending FQHE and Tonks-Girardeau ideas with Rashba SO coupling presents one with a number of challenges.  Indeed, the standard way
of introducing Chern-Simons transformation\cite{Jain, Halperin}
  essentially relies on the spinless nature of  particles.  Here
we suggest a way to generalize it for spinor particles with strong SO coupling. It achieves the goal of fermionization  of bosons, but fails to eliminate completely the interactions between the composite {\em spinless} fermions. Still the fermionic interaction energy can be parametrically smaller than the bosonic one. The crucial observation here is that the residual fermion-fermion interactions appear to be proportional to the angle between  momenta of two scattering particles. Therefore, confining the Fermi sea of composite fermions to a small fraction of the momentum space, allows one to
lower their interaction energy. Similar ideas were recently put forward by Berg, Rudner and Kivelson (BRK)\cite{berg} in the context of
Fermi gas with Rashba SO coupling. They called the resulting time-reversal symmetry broken state a nematic. Here we adopt their construction for the {\em composite} fermions.

The paper is organized as follows: in section \ref{sec:rashba} we introduce Rashba SO coupling on a single-particle level. We also discuss
earlier mean-field theories for bosonic many-body groundstates along with BRK nematic state for  many-body fermionic system.  In section
\ref{sec:comp-fermions} we introduce fermionization of spinor bosons and evaluate the energy of the resulting composite fermion
state. In section \ref{sec:discussion} we discuss the results: present an emerging groundstate phase diagram of Rashba bosons, list possible experimental consequences and generalizations of our approach. Some technical details of the calculations are relegated to the Appendix.

%%%%%%%%%%%%%%%%%%%%%%%%%%%%%%%%%%%%%%%%%%%%%%%%%%%%%%%%%%%%%%%%
%\begin{figure}[t]
%\centerline{\includegraphics[width=75mm,angle=0,clip]{Phase.eps}}
%\caption{(Color online) }
% \label{MC}
%\end{figure}
%%%%%%%%%%%%%%%%%%%%%%%%%%%%%%%%%%%%%%%%%%%%%%%%%%%%%%%%%%%%%%%%%

%%%%%%%%%%%%%%%%%%%%%%%%%%%%%%%%%%%%%%%%%%%%%%%%%%%%%%%%%%%%%%%%%%%%%%%%%%%%%%

%\section{Cold atoms with Spin-Orbit interaction  in two dimensions }
\section{Rashba spin-orbit coupling}
\label{sec:rashba}

\subsection{Single particle spectrum}
\label{sec:single-particle}

The single-particle Hamiltonian with the Rashba spin-orbit coupling in two dimensions takes the form
\bea
\label{H-single-particle}
H_0= -\frac{{\bm \nabla}_{\bf r}^2}{2 m}+ i v \hat{\bf z}\cdot[\bm{\sigma}\times {\bm \nabla_{\bf r}}]\,,
\eea
where $v$ is spin-orbit coupling constant having dimensionality of velocity, ${\bm \nabla}_{\bf r}=(\partial_x,\partial_y)$ and $\bm{\sigma} =(\sigma_x,\sigma_y)$ is  vector of Pauli
matrices acting on two component spinor $\psi(\bf r)=\left(\psi(\bf r,\uparrow),\psi(\bf{r},\downarrow)\right)$. In 2D the Rashba term in Eq.~(\ref{H-single-particle}) may be transformed to another form $i v \bm{\sigma}\cdot {\bm \nabla_r}$  by $\pi/2$ rotation
in the spin space.

In addition to translational and rotational symmetries the Hamiltonian $H_0$ commutes with the two discrete $Z_2$ symmetry operations: time-reversal $\hat T$ and 2D parity $\hat P$. They may be defined in the following way
\be
                              \label{T}
\hat T \left( \begin{array}{r}
\psi(\bf r,\uparrow)\\\psi(\bf r,\downarrow)
\end{array} \right) =
\left( \begin{array}{r}
\bar\psi(\bf r,\downarrow)\\-\bar \psi(\bf{r},\uparrow)
\end{array} \right)
\ee
and
\be
                              \label{P}
\hat P \left( \begin{array}{r}
\psi(z,\uparrow)\\\psi(z,\downarrow)
\end{array} \right) =
\left( \begin{array}{r}
-i\psi(\bar z,\downarrow)\\i\psi(\bar z,\uparrow)
\end{array} \right),
\ee
where bar stands for complex conjugated and we introduced the complex 2D coordinate as $z=x+iy$. Notice that the parity operation in 2D is defined as $y\to -y$ and $x\to x$ and $\sigma_y$ multiplication. Indeed, reflection of both coordinates $x$ and $y$ is equivalent to $\pi$ rotation.
Of course, one could as well define parity as $x$ reflection and $\sigma_x$ multiplication. It is equivalent to the $P$ operation (\ref{P}) combined with the $\pi$ rotation.

\begin{figure}[t]
\centerline{\includegraphics[width=75mm,angle=0,clip]{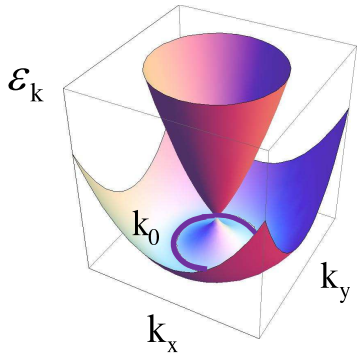}}
\caption{(Color online) Dispersion relation (\ref{sp1}) of particles with Rashba SO coupling. The degenerate groundstate
${\bf k}= k_0=mv$ is shown by full circle.     }
 \label{sp}
\end{figure}

Diagonalization of the Hamiltonian~(\ref{H-single-particle})  yields the following
single-particle spectrum:
\bea
                                                                \label{sp1}
\varepsilon_{{\bf k},\gamma}= \frac{{k}^2}{2 m}+ \gamma v k.
\eea
Here index $\gamma =\pm 1$ labels the two brunches of the spectrum depicted in Fig.~\ref{sp}.
The corresponding eigenfunctions are plane waves  in coordinate space  multiplying coordinate-independent spinor,
whose form depends, however, on the momentum ${\bf k}$ as
\bea
\label{wave1}
\psi_{{\bf k},\gamma}({\bf r},s)=\frac{1}{\sqrt{2V}}
\left(
                                                           \begin{array}{c}
                                                             -i \gamma\, e^{-i \text{arg} ({\bf k})} \\
                                                             1 \\
                                                           \end{array}
                                                         \right)_s\, e^{i {\bf k} {\bf r}}.
\eea
where $ \text{arg} ({\bf k})$ is the angle between the momentum vector ${\bf k}$ and the $k_x$-axis, $V$ is the system's volume.
The spin  is directed in the $x-y$ plane and is rotated relative to the momentum  direction by $\gamma\pi/2$. Hereafter we consider SO energy scale $\epsilon_0=mv^2/2$ as the largest energy in the problem.
One thus expects the groundstate of an interacting system to be confined to the part of the Hilbert space projected on the lower branch $\gamma=-1$.
Clearly
the two spin components of all such states are connected by the  following unitary transformation
$\psi_{{\bf k},-}({\bf r},\uparrow)= i e^{-i \arg({\bf k})}\psi_{{\bf k},-}({\bf r},\downarrow)$. For a generic wavefunction belonging to the lower branch  the relation between its up and down components   acquires the form
\bea
\label{kernel}
\psi({\bf r}, \uparrow)&=&\int d {\bf r}^\prime\,  {\cal R}({\bf r}-{\bf r}^\prime)\, \psi({\bf r}^\prime,\downarrow),
\eea
where the spin-raising kernel
\bea
\label{kernel1}
{\cal R}({\bf r}-{\bf r}^\prime)&=&-\frac{1}{2\pi }\, \frac{e^{-i \mathrm{arg}({\bf r}-{\bf r}^\prime)}}{ ({\bf r}-{\bf r}^\prime)^2}
\eea
is the Fourier transform of $ie^{-i \arg({\bf k})}$.
Importantly, the ${\cal R}$-operator in real space has the unitarity property, $\int d{\bf r} \bar {\cal R}({\bf r}-{\bf r}_1){\cal R}({\bf r}-{\bf r}_2)=\delta({\bf r}_1-{\bf r}_2)$,
which is the direct consequence of the unitarity of the spin-rising transformation in the momentum space representation.

The most notable feature of the dispersion relation (\ref{sp1}) is that its groundstate is degenerate along the
circle in the momentum space $k=k_0=mv$. As a result the many-body groundstate of $N$ {\em non-interacting } bosons is highly degenerate
(not so for fermions, though). Indeed, any occupation of the states along the groundstate   circle $k=k_0$ yields exactly the same
kinetic energy $-Nk_0^2/2m$. Hereafter we shall measure the energy from that value, taking it for the origin of the energy axis. It is therefore only the interactions, which may break the degeneracy and select a true groundstate. The situation is similar to
a partially filled Landau level in the context of FQHE. There too the kinetic energy is fully degenerate and
the groundstate is solely determined by the interparticle interactions.

%%%%%%%%%%%%%%%%%%%%%%%%%%%%%%%%%%%%%%%%%%%%%%%%%%%%%%%%%%%%%%%
%%%%%%%%%%%%%%%%%%%%%%%%%%%%%%%%%%%%%%%%%%%%%%%%%%%%%%%%%%%%%%%%%

\subsection{Bosons with Rashba dispersion}
\label{sec:bosons}

Let us consider $N$ particles with the $s$-wave short-range interactions of the form
\bea
                                                     \label{Hint}
H_\mathrm{int} =  \frac{1}{2m}
\sum_{i,j}^N \delta({\bf r}_i - {\bf r}_j)\big[g_0 + g_2 \sigma_z^{(i)}\sigma_z^{(j)} \big],
\eea
where  $g_0$ is the spin-isotropic dimensionless interaction constant, while $g_2$ is the spin-anisotropic
interaction constant.

One can now evaluate interaction energy of certain simple $N$-body Bose states.
One such state is a Bose-Einstein condensate in a single state belonging to a degenerate manifold of the single-particle ground states, i.e. a state
with momentum ${\bf k}$, such that $k=k_0$ and $\gamma=-1$,
\bea
                                              \label{cond}
\Psi^{(0)}_B=\prod_{i=1}^N \psi_{{\bf k},-}({\bf r}_i,s_i)\,.
\eea
This wavefunction is obviously symmetric with respect to the permutation of any two pairs $({\bf r}_i,s_i)\leftrightarrow ({\bf r}_j,s_j)$.
Such a state is not symmetric under time-reversal transformation $\hat T$, Eq.~(\ref{T}), because of unequal population of ${\bf k}$ and $-{\bf k}$ states. We shall call it thus  time-reversal symmetry broken (TRSB) state. Note that this state does {\em not} break the parity $\hat P$, Eq.~(\ref{P}). To see it most clearly one may choose momentum ${\bf k}$ direction to be along the $x$-axis.
Calculating the expectation value of the interaction energy (\ref{Hint}) over TRSB state, one finds
\bea
                                               \label{cond-energy}
E^{(0)}_\mathrm{int} =  \frac{N^2 }{2mV}\, g_0 \, .
%{ is correct}
\eea
Since the kinetic energy in the state (\ref{cond}) is zero, the interaction energy (\ref{cond-energy})  coincides with the total one.

One may consider now a Bose condensate built on a coherent superposition of say two states ${\bf k}_1$ and ${\bf k}_2$ both belonging to
the degenerate manifold $k=k_0$:
\bea
                                              \label{phi-cond}
\Psi^\mathrm{(\phi)}_B=\prod_{i=1}^N \frac{1}{\sqrt{2}}\left[  \psi_{{\bf k}_1,-}({\bf r}_i,s_i)+  \psi_{{\bf k}_2,-}({\bf r}_i,s_i)
\right]  \,.
\eea
The corresponding interaction (and thus total) energy is found to be
\bea
                                               \label{phi-cond-energy}
E^{(\phi)}_\mathrm{int} =  \frac{ N^2 }{2mV}\, \left[g_0+ {g_0\over 2}\cos^2 {\phi\over 2} + {g_2\over 2} \sin^2 {\phi\over 2} \right] \, ,
\eea
where $\phi = \text{arg} ({\bf k}_1) - \text{arg} ({\bf k}_2)$ is the angle between the two states of the degenerate
manifold. It is clear that, provided the spin-isotropic interaction is repulsive $g_0>0$, the only way the state (\ref{phi-cond})
may be energetically more favorable than the state (\ref{cond}) is if $g_2<0$. In the latter case the most favorable choice
of the two states corresponds to $\phi =\pi$, i.e. ${\bf k}_2=-{\bf k}_1$,  with the interaction energy $E^{(\pi)}_\mathrm{int} = ( N^2 /2mV)\, \left[g_0+ g_2/2\right] $.
Such a state represents  a spin density wave (SDW) with a uniform total density and the two spin components oscillating harmonically
out of phase. It is symmetric with respect to both time-reversal and parity symmetries, Eqs.~(\ref{T}), (\ref{P}), but breaks the rotational symmetry.
In either of these states the kinetic energy is zero.

It was conjectured \cite{zhai,Wu} that TRSB  $\Psi^{(0)}_B$, (\ref{cond}),  and the SDW state $\Psi^{(\pi)}_B$, (\ref{phi-cond}), are the many-body groundstates of the Rasba interacting bosons for $g_2>0$ and $g_2<0$ correspondingly. It was later
suggested \cite{Lamacraft} that the transition between TRSB and SDW  states is shifted towards positive $g_2$ due to
a ceratin admixture of coherently occupied BCS-like pairs of ${\bf k}$ and $-{\bf k}$ states.  We notice that for both of these states the chemical potential scales linearly with the density $n=N/V$,
\be
                                                   \label{mu-B}
\mu_B=\partial E_\mathrm{int}/\partial N  \propto n\,.
\ee
It is instructive to compare this scaling of the bosonic chemical potential with the corresponding fermionic one.

%%%%%%%%%%%%%%%%%%%%%%%%%%%%%%%%%%%%%%%%%%%%%%%%%%%%%%%%%%%%%%%%%%%%%%%%%%%%%%

\subsection{Fermions with Rashba dispersion}
\label{sec:fermions}

Unlike bosons, the {\em non-interacting} Fermi gas exhibits the unique ground-state. It is given by the rotationally symmetric Fermi sea with the Fermi surface
consisting of two concentric circles with the radii $k_{F\pm}= k_0 \pm 2\pi n/k_0$, hereafter the small density $n\ll k_0^2$ is assumed.
The corresponding Fermi energy measured from the bottom of the spectrum is
\bea
                                             \label{Fermi-energy}
\mu_F^{(0)}=\frac{ (\pi n)^2}{2m k_0^2}\,.
\eea
Notice the fact that $\mu_F^{(0)} \propto n^2$, which is usually the feature of 1D Fermi gas. Here it happens because the single-particle density of states exhibits 1D-like behavior $\nu(\epsilon) \propto \epsilon^{-1/2}$ close to the bottom of the Rashba circle.
The interesting observation is that at small enough density $n\lesssim g_0k_0^2$  the
chemical potential of interacting bosons appears to be bigger than that of the free Fermi gas with the same dispersion relation (\ref{sp1}). Before making conclusions from this observation one  needs to consider the interaction energy (\ref{Hint}) of the Fermi sea
state.

Spinless (or fully spin-polarized) fermions do not interact through short-range interactions. In our case the particles have spin, which is locked to their orbital momenta. As a result the symmetric Fermi sea, described above, contains all spin directions in {\em x-y} plane with equal weights. Since two fermions with opposite spins interact through the short range interaction (\ref{Hint}), one expects the average
interaction energy of the symmetric Fermi sea to be of the same order as bosonic one $\propto g_0N^2/mV$ and thus $\mu_F\propto n$ similarly to the
bosonic condensates. It was recently noticed by Berg, Rudner and Kivelson  \cite{berg} that low density Rashba fermions may have
parametrically lower groundstate energy if they form  a {\em nematic} state.

To motivate the idea, let us consider two-fermion state as a Slater determinant built on single particle states (\ref{wave1}) with momenta ${\bf k}_{1,2}$ close to the Rashba circle $\Psi_F=[\psi_{{\bf k}_1,-}({\bf r}_1,s_1)\psi_{{\bf k}_2,-}({\bf r}_2,s_2)-
\psi_{{\bf k}_1,-}({\bf r}_2,s_2)\psi_{{\bf k}_2,-}({\bf r}_1,s_1)]/\sqrt{2}$. The interaction energy (\ref{Hint}) of such state is given by
$E_\mathrm{int}=\sin^2(\phi/2)(g_0+g_2)/2m$, where $\phi=\mathrm{arg}({\bf k}_1) -  \mathrm{arg}({\bf k}_2)$. Therefore  the interaction energy between fermions tends to zero if $\mathrm{arg}({\bf k}_1) \to  \mathrm{arg}({\bf k}_2)$, essentially because the spins are aligned in this limit and
fermions with the same spin can not interact through the short-range interaction potential. The BRK nematic state takes advantage of this observation.

\begin{figure}[h]
\centerline{\includegraphics[width=75mm,angle=0,clip]{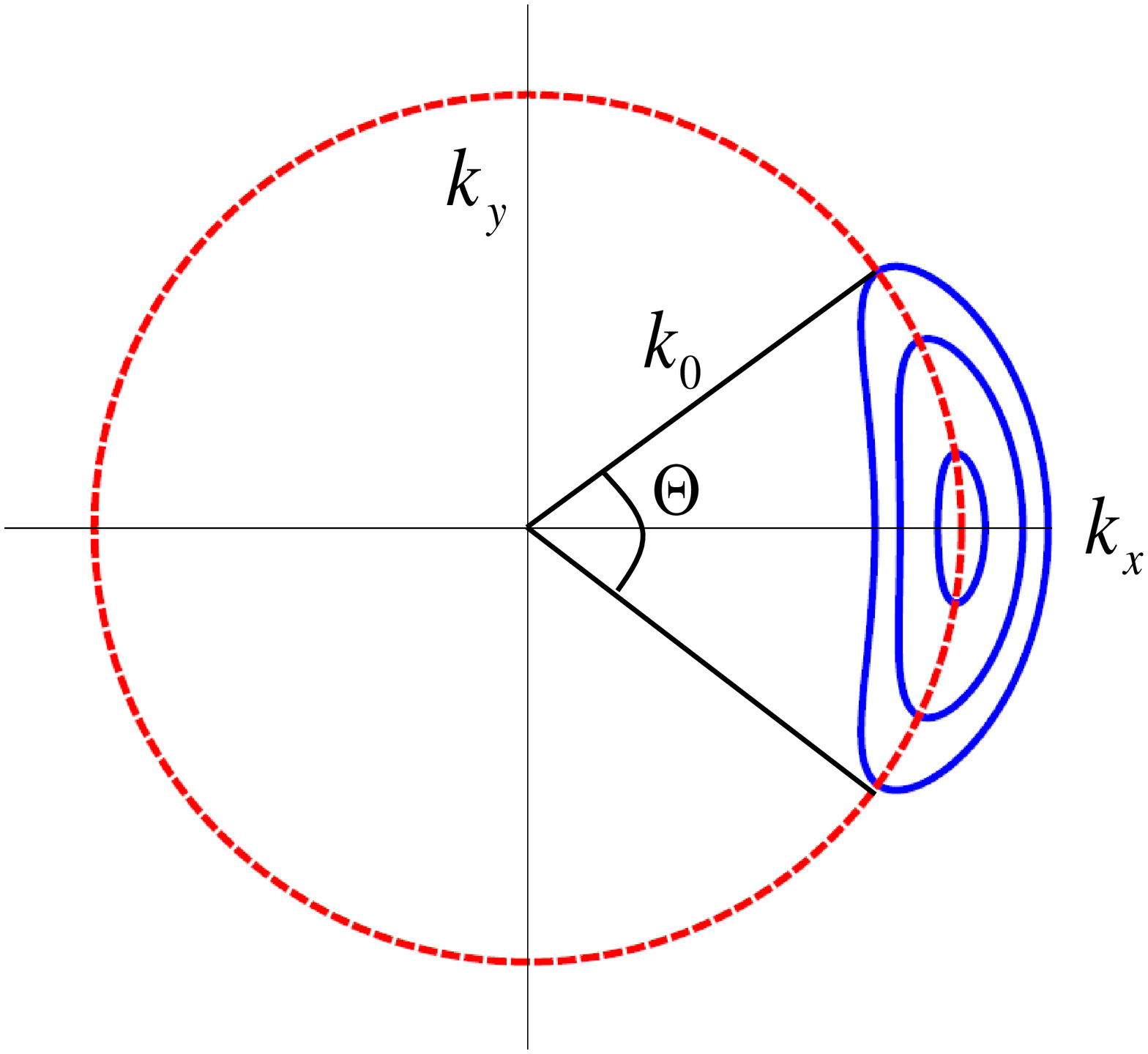}}
\caption{(Color online) Hartree-Fock Fermi surfaces of  BRK nematic states with three different densities. The absolute value of the momenta are restricted to the vicinity  of the Rashba circle ${\bf k}=k_0$, while their angular directions are restricted to the angle $\Theta \propto [n/k_0^2(g_0+g_2)]^{1/4}$.}
 \label{fs}
\end{figure}

To describe such a state
qualitatively let us imagine that the many-body fermionic state $\Psi_F$ is constructed  as a Slater determinant of states with
the angular directions confined to the angular segment of the momentum (and spin) space of size $\Theta\ll 2\pi$ and momenta close to the spin-orbit circle $||{\bf k}_j|- k_0|<k_F^{(\Theta)}$, Fig.~\ref{fs}.  The corresponding Fermi momentum is found from the condition $2k_F^{(\Theta)} k_0\Theta=(2\pi)^2 n$. As a result the corresponding kinetic energy per particle  $E_\mathrm{kin}/N \propto [k_F^{(\Theta)}]^2/m\propto n^2/(m\Theta^2k_0^2)$. On the other hand,
the interaction energy per particle is $E_\mathrm{int}/N \propto (g_0+g_2)n\Theta^2/m$, with the factor $\Theta^2$ originating from $\sin^2(\phi/2)\sim \Theta^2$. One can now minimize the sum of kinetic and interaction energy over $\Theta$ to find $\Theta \propto [n/k_0^2(g_0+g_2)]^{1/4}$, which is indeed small as long as
\be
                              \label{density-small}
n \ll n_0 = k_0^2(g_0+g_2)\, .
\ee
With this $\Theta$ the chemical potential of  short-range {\em interacting} fermions with Rashba spin-orbit coupling is found to be $\mu_F \propto n^{3/2} \sqrt{g_0+g_2}/mk_0$, which is the result of BRK\cite{berg}.
In the small density limit (\ref{density-small}) the latter is bigger than that of non-interacting Rashba fermions (\ref{Fermi-energy}), but is
advantageous over bosonic TRSB and  SDW states (\ref{mu-B}): $\mu_F^{(0)} < \mu_F <\mu_B$. Notice also the non-analytic dependence
of $\mu_F$ on the interaction strength, indicating the non-perturbative nature of this result.
We show in Section \ref{sec:HF} that the composite fermion nematic state may be described quantitatively within Hartree-Fock self-consistent mean-field treatment, Fig.~\ref{fs}.

%%%%%%%%%%%%%%%%%%%%%%%%%%%%%%%%%%%%%%%%%%%%%%%%%%%%%%%%%%%%%%%
%%%%%%%%%%%%%%%%%%%%%%%%%%%%%%%%%%%%%%%%%%%%%%%%%%%%%%%%%%%%%%%%%

\subsection{Can bosons be fermions?}
\label{sec:can-bosons-be}

\begin{figure}[t]
\centerline{\includegraphics[width=75mm,angle=0,clip]{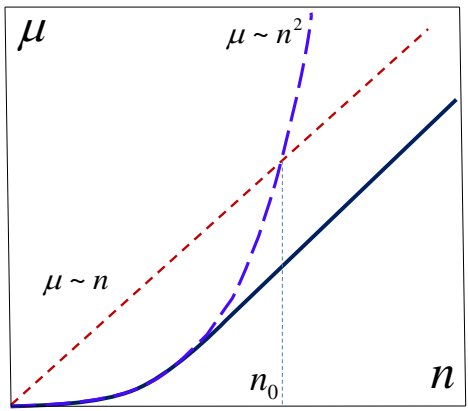}}
\caption{(Color online) Equations of state $\mu(n)$ of 1D quantum gases with short-range interactions: spinless Fermi gas (long-dashed); mean-field approximation for Bose gas (dashed); exact result\cite{LiebLiniger} for bosons (full). Notice that bosonic groundstate energy $E=\int \mu dn$ is {\em always} smaller than the fermionic one, although the mean-field treatment suggests otherwise for $n\lesssim n_0=g_0$.   }
 \label{fig-1D}
\end{figure}

We have arrived thus to the conclusion that at low density  (\ref{density-small}) the chemical potential and groundstate energy of fermionic many-body system  is smaller than the corresponding bosonic one. The similar situation {\em seemingly} happens  in a 1D system of spinless particles with short range interaction potential $(g_0/m)\delta(x_i-x_j)$. There too the mean-field treatment of bosons suggests that $\mu_B\sim g_0n$. On the the other hand, spinless fermions (which are not affected by short-range interactions due to the Pauli principle)
exhibit $\mu_F\sim n^2$. It thus seems that at a small enough density the fermionic groundstate energy is smaller than the bosonic one.
The actual situation is, of course, very different\cite{LiebLiniger}. At small density bosonic many-body groundstate wavefunction $\Psi_B$ approaches the {\em symmetrized} fermionic one
\be
                                              \label{boson-fermion-1D}
\Psi_B(x_1,\ldots, x_N) = \prod_{i,j}^N\mbox{sign}(x_i-x_j) \Psi_F(x_1,\ldots, x_N)\,,
\ee
where $\Psi_F$ is the fermionic Slater determinant occupying a finite portion of the momentum space $-\pi n\leq k\leq \pi n$.
It is important to notice that, although $\mbox{sign}(x_i-x_j)$ is undefined at $x_i=x_j$, the fermionic part cancels at all such points
making $ \Psi_B(x_1,\ldots, x_N)$ well-defined in the entire space.  Due to the same observation there is no interaction energy cost for the short-range repulsion. The corresponding kinetic energy per particle $\propto (\pi n)^2/2m$  is small in the
low density limit. This is the so-called Tonks-Girardeau limit~\cite{tonks,girardeau,yang}, where bosons redress themselves as fermions.  This allows them to take  advantage of the wavefunction, which is nullified at all points where any two particles approach each other. As a result they avoid paying
short-range interaction energy cost, while corresponding kinetic energy cost appears to be worth the bargain at low density. The corresponding
1D equations of state are depicted in Fig.~\ref{fig-1D}.
%%%%%%%%%%%%%%%%%%%%%%%%%%%%%%%%%%%%%%%%%%%%%%%%%%%%%%%%%%%%%%%
%%%%%%%%%%%%%%%%%%%%%%%%%%%%%%%%%%%%%%%%%%%%%%%%%%%%%%%%%%%%%%%%%

It is thus tempting to speculate that our spinfull 2D system may benefit from a similar construction. Namely, at low density bosons may want to redress themselves as fermions to take  advantage of their lower interaction energy. The blueprints of the boson-fermion correspondence in 2D
are provided by FQHE studies \cite{Jain,Halperin}, where the correspondence is achieved by ascribing Chern-Simons phase factor to many-body wavefunctions. Such factor takes care of the proper symmetry  of  wavefunctions. It comes with the price however: the gauge magnetic field which has a form of delta-functional flux tubes attached to every particle. It is believed that (under proper conditions) this magnetic field may be substituted by a uniform one, making the problem analytically tractable. In the next section we develop
a similar strategy for the chiral spinfull boson-fermion correspondence.

\section{Composite fermion state}
\label{sec:comp-fermions}

\subsection{Na\"ive fermionization attempt}
\label{sec:naive}

Our goal is to construct  a variational groundstate for Rashba bosons based on the composite
fermion idea. We argue here that in the small density limit such a state, inspired by the Tonks-Girardeau limit (\ref{boson-fermion-1D}),
is advantageous over both TRSB (\ref{cond}) and SDW  (\ref{phi-cond}) bosonic states.
The two main differences with the Tonks-Girardeau case are: (a) the 2D nature of our problem and (b) the spinfull nature of the particles.
It is still tempting to straightforwardly generalize Eq.~(\ref{boson-fermion-1D}) to the 2D spinfull case by taking the variational ground state of the Bose gas in the following form:
\bea
\label{CB-1}
&&\Psi_B({\bf r}_1,s_1, \cdots {\bf r}_N,s_N)\\
&&= \prod_{i<j}e^{i \arg({\bf r}_i-{\bf r}_j)}
\Psi_F({\bf r}_1,s_1, \cdots {\bf r}_N,s_N),  \nn
\eea
where $\Psi_F({\bf r}_1,s_1, \cdots {\bf r}_N,s_N)$ is an N-particle fermionic wavefunction. The Chern-Simons phase
$e^{i \arg({\bf r}_i-{\bf r}_j)} = {(z_i-z_j)}/{|z_i-z_j|}$, with $z_j$ labeling complex spatial coordinates, $z_j=x_j+i y_j$, is  antisymmetric
with respect to exchange of any two coordinates. As a result,
many-body wavefunction, $\Psi_B({\bf r}_1,s_1, \cdots {\bf r}_N,s_N)$,
is symmetric under permutation of pairs ${\bf r}_i, s_i$ and ${\bf r}_j, s_j$ of coordinates and spins,
$s_i=\uparrow,\downarrow$, if the fermionic wave function, $\Psi_F({\bf r}_1,s_1, \cdots {\bf r}_N,s_N)$,
is antisymmetric with respect to the same permutations. One way of writing the latter is to take it as $N\times N$ Slater
determinant of e.g. single-particle spinor wave functions, $\psi_{{\bf k}_j,-}({\bf r}_i,s_i)$, Eq.~(\ref{wave1}), where $i,j=1,2, \ldots, N$.

While being probably the most straightforward way of guessing a spinfull composite fermion state, Eq.~(\ref{CB-1}) has a number of
fatal drawbacks. First, one might na\"ively expect that similarly to Eq.~(\ref{boson-fermion-1D}), this ansatz maps the interacting bosonic system onto a system of non-interacting fermions. A closer look shows that this is not the case. Indeed, although the Slater determinant implies that the fermionic wave function has zeros at coinciding spatial points {\em and spins}, however for coinciding points and {\em different spins} the wavefunction is not nullified. As a result the composite fermions with opposite spins still do interact, despite the short-range nature of the interaction potential.

Even more serious problem with the trial wavefunction (\ref{CB-1}) is that it is not well-defined for coinciding points and opposite spins.
This is due to the fact that the Chern-Simons phase is singular at coinciding points, while the fermionic part $\Psi_F$ is not vanishing if spins are opposite. This ambiguity leads to logarithmic divergent contributions to the average kinetic energy of the  state (\ref{CB-1}). Indeed, consider
the part of the kinetic energy $(2m)^{-1}\int d{\bf r_i} |\nabla_{{\bf r}_i} \Psi_B|^2$, where the gradient operators act on the Chern-Simons phases. This leads to the kinetic energy contribution of the form
\be
                                                                                            \label{log-singular}
\frac{1}{2m}\!\! \int\!\! d{\bf r_i}\! \sum_{j,j'} \frac{({\bf r}_i-{\bf r}_j)\cdot ({\bf r}_i-{\bf r}_{j'}) }{|{\bf r}_i-{\bf r}_j|^2 |{\bf r}_i-{\bf r}_{j'}|^2}   \,  \big|\Psi_F({\bf r}_1,s_1, \cdots {\bf r}_N,s_N)\big|^2.
\ee
The diagonal terms $j=j'$ in this double sum lead to the integrals of the form $\int d{\bf r_i}/|{\bf r}_i-{\bf r}_j|^2$, which  exhibit logarithmic behavior when ${\bf r}_i\approx {\bf r}_j$.
If particles $i$ and $j$ have the same spin, $\Psi_F=0$ at ${\bf r}_i={\bf r}_j$ cutting the logarithmic divergence of the integral at small distances $\sim |{\bf k}_i-{\bf k}_j|^{-1}$ (at large distances it is cut at a typical
interparticle distance $n^{-1/2}$ due to the random sign of the numerator in Eq.~(\ref{log-singular})). However for opposite spins  $\Psi_F\neq 0$ at ${\bf r}_i={\bf r}_j$ and the integral (\ref{log-singular}) diverges in all such points.
The result is logarithmical divergent chemical potential, making the trial wavefunction (\ref{CB-1}) essentially useless.
One should thus look for an alternative way to introduce composite fermion state for spinfull Rashba particles, which avoids logarithmic divergent terms in the kinetic energy.

\subsection{Fermionization}
\label{sec:ferm}

The idea for an alternative scheme comes from the observation that at small density all relevant energy scales are much smaller
than  SO energy $\epsilon_0=mv^2/2$. Therefore one would like to have a many-body state which is projected onto the
Hilbert subspace of the lower spin-orbit branch $\gamma=-1$, Eq.~(\ref{sp1}). On the single particle level such a projection
is achieved by ensuring the relation (\ref{kernel}) between up and down components of the spinor. One can straightforwardly generalize it
for a many-body wavefunction. To this end one should specify a fully symmetric in the coordinate space wavefunction of the minimal spin
\be
                                                                \label{alldown}
\Psi_{\downarrow\ldots\downarrow}({\bf r}_1,\ldots, {\bf r}_N) =
\Psi({\bf r}_1\downarrow,\ldots, {\bf r}_N\downarrow)\,.
\ee
Then all other spin components of the fully projected wavefunction may be uniquely determined from the minimal spin component by successive
application of the spin raising non-local operator ${\cal R}$, Eq.~(\ref{kernel1}), e.g.
\bea
                                                                \label{someup}
\Psi({\bf r}_1\!\uparrow,\ldots, {\bf r}_N\!\downarrow) =&&\! \int\!\! d {\bf r}_1^\prime {\cal R}({\bf r}_1-{\bf r}_1^\prime)\Psi_{\downarrow\ldots\downarrow}({\bf r}_1^\prime,\ldots, {\bf r}_N)\,,\nn \\
\Psi({\bf r}_1\!\uparrow,\ldots, {\bf r}_N\!\uparrow) =&&\! \int \!\! d {\bf r}_1^\prime  \ldots d{\bf r}_N^\prime {\cal R}({\bf r}_1-{\bf r}_1^\prime\!)\ldots {\cal R}({\bf r}_N-{\bf r}_N^\prime\!) \nn \\
&& \quad\quad \Psi_{\downarrow\ldots\downarrow}({\bf r}_1^\prime,\ldots, {\bf r}_N^\prime)\,.
\eea
It is easy to see that all the components defined this way are symmetric with respect to simultaneous interchange of
${\bf r}_i,s_i$ and  ${\bf r}_j,s_j$.

One can now fermionize such a wavefunction by writing the spatially symmetric minimal spin component (\ref{alldown}) as a product of Chern-Simons
phase and fully antisymmetric {\em spinless} fermionic wavefunction. The latter will be shown to describe the anisotropic nematic state. It is
therefore convenient to incorporate the same anisotropy in the Chern-Simons phase too. We thus define rescaled coordinates $\tilde x=\alpha x$, $\tilde y=y/\alpha$ and  $\tilde{\bf r} =(\tilde x,\tilde y)$, where the density-dependent scaling parameter $\alpha$ will be specified in Section \ref{sec:CS}.
The fermionized minimal spin component is then written as:
\be
\label{CB-11}
\Psi_{\downarrow\ldots\downarrow}({\bf r}_1, \ldots, {\bf r}_N)\! = \! \frac{1}{\sqrt{2^N}}\prod_{i<j}e^{i\lambda \arg(\tilde {\bf r}_i-\tilde {\bf r}_j)}
\Psi_F({\bf r}_1, \ldots, {\bf r}_N)\, ,
\ee
where we have introduced chirality factor $\lambda=\pm 1$, which defines
the direction of the Chern-Simons flux relative to the spin chirality. The many-body fermionic wavefunction $\Psi_F({\bf r}_1, \ldots, {\bf r}_N)$
is antisymmetric with respect to  permutation of any two of its  spatial arguments. Thanks to the Chern-Simons phase factor the left hand side
of Eq.~(\ref{CB-11}) is fully symmetric both in coordinate and spin spaces.
Notice that, unlike the earlier attempt, Eq.~(\ref{CB-1}),
the wavefunction (\ref{CB-11}) is everywhere well-defined. This is the case because the Chern-Simons phase multiplies the fully antisymmetric function of {\em spinless} fermions, which cancels if any of its spatial arguments coincide. This became possible by adding the Chern-Simons factor to a fully spin-symmetric
minimal spin component only. The higher spin components are then built up from the fermionized minimal spin bosonic function (\ref{CB-11}) by a successive application of the spin raising kernel, Eq.~(\ref{someup}).  As a result {\em all} $2^N$ spin components of the $N$-body bosonic
wavefunction (\ref{someup}), (\ref{CB-11}) are well-defined functions in the entire $2N$-dimensional coordinate space. This is unlike the earlier attempt, see Eq. (15), which resulted in the logarithmic divergences in the kinetic energy.

Of course, there is a complimentary construction where one fermionizes the maximal spin component   $\Psi_{\uparrow\ldots\uparrow}({\bf r}_1, \ldots, {\bf r}_N)$ and builds the lower spin components by successive action of spin lowering operator. These two states are related by the parity transformation (\ref{P}) acting on the coordinates and spins of all $N$ particles. Namely, acting with such parity operator on
the state (\ref{someup}), (\ref{CB-11}) built from the minimal spin with chirality $\lambda$, one obtains the state built from the maximal spin component with chirality $-\lambda$ and parity transformed, $y_j\to -y_j$,  fermionic wavefunction $\Psi_F$. Indeed, the parity operation (\ref{P}) interchanges spins and transforms $z_j\to \bar z_j$ resulting in $\lambda\to -\lambda$ correspondence. Since the parity operator $\hat P$ commutes with the full Hamiltonian $H_0+H_\mathrm{int}$, these two states are degenerate.
%By choosing one of them, the system spontaneously breaks the parity symmetry.

Let us now discuss the average kinetic energy of the state with the wavefunction  (\ref{alldown}), (\ref{someup}). Evaluating the expectation value of the single-particle operator
incorporating  kinetic and spin-orbit parts (\ref{H-single-particle}), one finds
\bea
E_\mathrm{kin}= &&\,\, 2^N\!\! \int \!\! \prod\limits_{j=1}^N d {\bf r}_j  d{\bf r}_j^\prime \,
\bar\Psi_{\downarrow\ldots\downarrow}({\bf r}_1,\ldots, {\bf r}_N)
\sum\limits_{j=1}^N  \hat K({\bf r}_j - {\bf r}_j^\prime)\nn \\
&&\quad\quad\quad\quad\quad \times \Psi_{\downarrow\ldots\downarrow}({\bf r}_1^\prime,\ldots, {\bf r}_N^\prime)\,,
                                                            \label{kin-energy}
\eea
where the non-local operator of the kinetic energy acting on the minimal spin component is
\be
                                                                  \label{kin-kernel}
{\hat K}({\bf r} - {\bf r}^\prime)=
-\frac{\delta_{{\bf r} - {\bf r}^\prime} \nabla_{{\bf r}^\prime}^2 }{2 m} - \frac{ v}{2}
\left[\bar{\cal R}({\bf r'}\!-\!{\bf r})\partial^-_{\bf r'}
- \partial^+_{\bf r} {\cal R}({\bf r}\!-\!{\bf r'})\right] ,
\ee
here we employed  notations $\partial^\pm_{\bf r}=\partial_{x}\pm i\partial_{y}$.
This form is obtained by expressing the spin-up states  in terms of the spin-down states
with the help of  Eq.~(\ref{kernel}) and using unitarity of the spin-raising operator.
The Fourier transformation of the kernel (\ref{kin-kernel}) results in the lower branch, $\gamma=-1$, of the  single particle spectrum Eq.~(\ref{sp1})
$\hat K({\bf k})={\bf k}^2/2 m- v |{\bf k}|=\varepsilon_{k,-}$, which is, of course, expected for the projected wavefunction in the form (\ref{alldown}), (\ref{someup}). It is exactly the reason to build the higher spin states with the help of the spin-raising operator (\ref{someup})
to ensure that the kinetic energy belongs to the lower spin-orbit branch.

The  non-analytic behavior of the spectrum (\ref{sp1}) at
$k=0$ translates  into the non-local behavior of the kinetic energy kernel (\ref{kin-kernel}) in the coordinate representation. This non-locality complicates the way the kinetic energy operator acts on the Chern-Simons phase in Eq.~(\ref{CB-11}). On the other hand,
the low energy part of the Hilbert space is located close to the Rashba circle $|{\bf k}|= k_0$, i.e. far away from the $k=0$ singularity. Below we
discuss  variational choices for the many-body fermionic wavefunction $\Psi_F$, which explicitly include only momentum components localized around
$|{\bf k}|= k_0$ circle. For those components the non-analyticity at $k=0$ and thus non-locality of the kinetic energy in the coordinate
space are not essential.
Therefore the single-particle kinetic energy spectrum near its minimum may be approximated by an analytic function of momentum (and thus local differential operator in the coordinate representation) as, e.g.,
\bea
\label{appspec}
K({\bf k})=-\epsilon_0+(|{\bf k}|^2-k_0^2)^2/(8mk_0^2)\, .
\eea
Its action on Chern-Simons phase results into the substitution of the gradient operators by the covariant derivatives with the gauge vector potential (see below).

The important observation at this stage is that there are {\em no} logarithmical divergent contributions to the kinetic energy, which
were fatal for the earlier fermionization attempt (\ref{CB-1}). Indeed, the kinetic energy kernel (\ref{kin-energy}) acting on the Chern-Simons
phase (\ref{CB-11}) results, among others, in the term similar to (\ref{log-singular}). However, this time the fermionic wavefunction $\Psi_F$ does
not contain spin indices and cancels any time ${\bf r}_i={\bf r}_j$. As a result, all the logarithmic integrals are cut off at small distances
by a scale built into the fermionic wavefunction $\Psi_F({\bf r}_1,\ldots, {\bf r}_N)$.
%i.e. $|{\bf k}_i-{\bf k}_j|^{-1}$.
At large distances they are cut off at a typical interparticle distance.
%$n^{-1/2}$.
As we discuss in Section \ref{sec:CS},  these two length scales can be made of the same order by an appropriate choice of the rescaling parameter $\alpha$ in the Chern-Simons phase (\ref{CB-11}). This makes the logarithm to be a number of order one, showing that the localized flux-tube structure of the Chern-Simons magnetic field (see below) does not strongly affect the kinetic energy of the fermion state $\Psi_F$. It suggests the mean-field substitution of the flux tubes by a uniform magnetic field, analogous to those employed in FQHE literature, e.g. \cite{Jain,Halperin}.

Next we discuss inter-particle interactions, excluding for a while the Chern-Simons term in the wavefunction (\ref{CB-11}). Following  BRK, we employ Hartree-Fock approximation to minimize the total energy of interacting particles having spectrum (\ref{appspec}) and find that the chemical potential scales as $\mu\sim n^{3/2}$. Finally, we include the Chern-Simons term within the Hartree-Fock approximation for interactions. We argue that it does not affect the parametric dependence $\mu(n)\propto n^{3/2}$, although does affect the spectrum of single-particle excitations.

\subsection{Interactions}
\label{sec:int}

We focus now on the average energy of the short-range interactions (\ref{Hint}) in the many-body state specified by
Eqs.~(\ref{CB-11}) and (\ref{someup}). Since the minimal spin component (\ref{CB-11}) cancels at coinciding spatial points, it does
not contribute to the interaction energy. On the other hand, the higher spin components, built according to Eq.~(\ref{someup}), do not vanish at  coinciding points and thus lead to a non-zero interaction energy. At the first glance this observation makes the interaction energy of the
composite fermion state Eqs.~(\ref{CB-11}) and (\ref{someup}) to be of the same order as that of the bosonic condensate (\ref{cond}) or (\ref{phi-cond}), making   the entire construction useless. We show below that this is not the case thanks to the antisymmetric nature of
$\Psi_F({\bf r}_1,\ldots, {\bf r}_N)$. The latter leads to the Hartree {\em minus} Fock structure of the two-particle interactions, which nearly cancels the interaction
energy for particles with nearly collinear momenta (and thus spin) directions. On the other hand, the bosonic condensate  (\ref{cond}) admits only the Hartree contribution, while (\ref{phi-cond}) leads to Hartree {\em plus } Fock structure, not exhibiting the cancelation.

To derive the effective two-body interactions in the composite fermion state it is sufficient to consider
two particles in such state built of two single-particle states, e.g. $\psi_{{\bf k}_j}({\bf r})=e^{i{\bf k}_j \cdot {\bf r}}/\sqrt{V}$ with $j=1,2$ as
\bea
                                                          \label{two-particle}
\Psi_F({\bf r}_1,{\bf r}_2) &=& \frac{1}{\sqrt{2}}\,
\Big[ \psi_{{\bf k}_1}({\bf r}_1)  \psi_{{\bf k}_2}({\bf r}_2)    -
\psi_{{\bf k}_1}({\bf r}_2)  \psi_{{\bf k}_2}({\bf r}_1)  \Big], \nn\\
\label{2-particle1}
\Psi_{ \downarrow \downarrow}({\bf r}_1,{\bf r}_2)&=& \frac{1}{2}\, e^{i\lambda \arg(\tilde{\bf r}_1-\tilde{\bf r}_2)}\, \Psi_F({\bf r}_1,{\bf r}_2)
,\nn \\
\label{2-particle2}
\Psi_{\uparrow\downarrow}({\bf r}_1,{\bf r}_2)&=&\int d {\bf r}_1^\prime {\cal R}({\bf r}_1 - {\bf r}_1^\prime)\Psi_{ \downarrow \downarrow}({\bf r}_1^\prime,{\bf r}_2),\\
\Psi_{\downarrow\uparrow}({\bf r}_1,{\bf r}_2)&=&\int d {\bf r}_2^\prime  {\cal R} ({\bf r}_2 - {\bf r}_2^\prime)\Psi_{ \downarrow \downarrow}({\bf r}_1,{\bf r}_2^\prime),\nn\\
\Psi_{\uparrow\uparrow}({\bf r}_1,{\bf r}_2)&=&\int d {\bf r}_1^\prime d {\bf r}_2^\prime {\cal R}({\bf r}_1\! -\! {\bf r}_1^\prime)
{\cal R}({\bf r}_2\! - \!{\bf r}_2^\prime) \Psi_{ \downarrow \downarrow}({\bf r}_1^\prime,{\bf r}_2^\prime).\nn
\eea

The average interaction energy according to Eq.~(\ref{Hint}) is given by
\bea
\label{2-inter}
E_{\text{int}}&=&\frac{1}{2m}\int d{\bf r}\Big[(g_0+g_2) |\Psi_{ \uparrow \uparrow}^{(\lambda)}({\bf r},{\bf r})|^2  \\
&+& (g_0-g_2)|\Psi_{ \uparrow \downarrow}^{(\lambda)}({\bf r},{\bf r})|^2
+ (g_0-g_2)|\Psi_{ \downarrow \uparrow}^{(\lambda)}({\bf r},{\bf r})|^2   \Big] \nn
\eea
and therefore one needs to evaluate the wavefunction for higher spin components at coinciding spatial points.
%Notice that due to the bosonic symmetry of the wavefunction  $\Psi_{ \uparrow \downarrow}({\bf r},{\bf r})=\Psi_{ \downarrow \uparrow}({\bf r},{\bf r})$.
%The corresponding calculations are presented in the Appendix. Here we quote the results in the approximation where $|{\bf k}_1|=|{\bf k}_2|=k_0$:
They are given by:
\bea
                                                                        \label{coinciding-points}
\Psi_{ \uparrow \uparrow}({\bf r},{\bf r}) &=& \frac{1}{{2} V}\,\,e^{i{\bf r}({\bf k}_1 +{\bf k}_2)}\,
F_{{\bf k}_1,{\bf k}_2} \,, \\
%\Psi_{ \uparrow \downarrow}^{(+)}({\bf r},{\bf r})&=& 0  \,,
\quad \quad
\Psi_{ \uparrow \downarrow}({\bf r},{\bf r}) &=& \Psi_{ \downarrow \uparrow}({\bf r},{\bf r})= \frac{1}{{2} V}\,\, e^{i{\bf r}({\bf k}_1 +{\bf k}_2)}\,
G_{{\bf k}_1,{\bf k}_2}\,, \nn
\eea
where $F_{{\bf k}_1,{\bf k}_2}$ and $G_{{\bf k}_1,{\bf k}_2}$ are interaction form-factors evaluated in the Appendix. In the approximation where $|{\bf k}_1|\approx  |{\bf k}_2|\approx k_0$ and $\mathrm{arg}({\bf k}_{1,2}) \ll 2\pi$ we found:
\bea
                                                                    \label{form-factors}
F_{{\bf k}_1,{\bf k}_2} &=& ic\big({\mathrm{arg}({\bf k}_1)} -  {\mathrm{arg}({\bf k}_2)}\big) \,, \\
G_{{\bf k}_1,{\bf k}_2} &=& d \big({\mathrm{arg}({\bf k}_1)} -  {\mathrm{arg}({\bf k}_2)} \big) \,.\nn
\eea
Here $c=c_\lambda(\alpha)$ and $d=d_\lambda(\alpha)$ are numerical factors weakly dependent on CS chirality $\lambda$ and anisotropy $\alpha$. For example,  $|c_+(1)| \simeq 1.22$; $|c_-(1)| \simeq 0.85$; $c_\pm(0)=0$ and $d_+(1)=0$; $|d_-(1)| \simeq 1.41$; $|d_\pm(0)| \simeq  0.90$.
A very important observation is that the interaction energy tends to zero if $\mathrm{arg}({\bf k}_1) \to  \mathrm{arg}({\bf k}_2)$. The relative minus sign in Eqs.~(\ref{form-factors}) is entirely due to the composite fermion nature of the wavefunction  (\ref{2-particle1}). Indeed the two-particle
fermionic wavefunction (\ref{two-particle}) is zero if $\mathrm{arg}({\bf k}_1) =  \mathrm{arg}({\bf k}_2)$ due to Pauli blocking.
This offers
a possibility to construct a trial many-body wavefunction which takes advantage of the fact that the particles with nearly collinear spins (and thus momenta!) interact only weakly. As discussed in section \ref{sec:fermions}, the similar construction for fermions with the Rashba SO coupling was recently employed by BRK \cite{berg}, who showed that the optimal trial state is a {\em nematic} one.

\subsection{Hartree-Fock theory}
\label{sec:HF}

The secondary quantized interaction Hamiltonian for projected spinless fermions  takes the form
\be
                                                         \label{second-quantized}
\hat H_\mathrm{int} =
%&=&\sum\limits_{\bf k} \varepsilon_{{\bf k},-} c^\dagger_{\bf k} c_{\bf k}\\  &+&
\frac{1}{32 m} \!\!\! \sum\limits_{{\bf k}_1^\prime+{\bf k}_2^\prime={\bf k}_1+{\bf k}_2 }
\!\!\!\! M_{{\bf k}_1^\prime,{\bf k}_2^\prime}^{{\bf k}_1,{\bf k}_2} \,\,
c^\dagger_{{\bf k}_2^\prime} c^\dagger_{{\bf k}_1^\prime} c_{{\bf k}_1} c_{{\bf k}_2},
%\delta_{ {\bf k}_1^\prime + {\bf k}_2^\prime -{\bf k}_1 -{\bf k}_2  }
\ee
where the interaction matrix elements
\be
                                                                       \label{matrix-element}
M_{{\bf k}_1^\prime,{\bf k}_2^\prime}^{{\bf k}_1,{\bf k}_2} = (g_0+g_2)\bar F_{{\bf k}_1^\prime,{\bf k}_2^\prime} F_{{\bf k}_1,{\bf k}_2}
+2(g_0-g_2) \bar G_{{\bf k}_1^\prime,{\bf k}_2^\prime} G_{{\bf k}_1,{\bf k}_2} \,.
\ee
%where we  restricted ourselves with the positive chirality $\lambda=+1$ of the Chern-Simons phase and therefore
%the effective interactions originate entirely from  the up-up term in Eq.~(\ref{2-inter}). Notice that the interactions
depend on all four (two incoming and two outgoing) momenta and therefore interactions can't be reduced to the density-density  form.
Moreover in the coordinate representation such interactions  acquire essentially non-local $|{\bf r}-{\bf r}^\prime|^{-2}$ form, which
is a consequence of the projection on the lower SO branch \cite{altman}. In principle the projection, Eqs.~(\ref{alldown}) and (\ref{someup}),
generates also three- and more-particle interactions. One may check however that in the dilute limit (\ref{density-small}) their effect is
negligibly small in comparison with the two-particle term kept in Eq.~(\ref{second-quantized}).

\begin{figure}[t]
\centerline{\includegraphics[width=65mm,angle=0,clip]{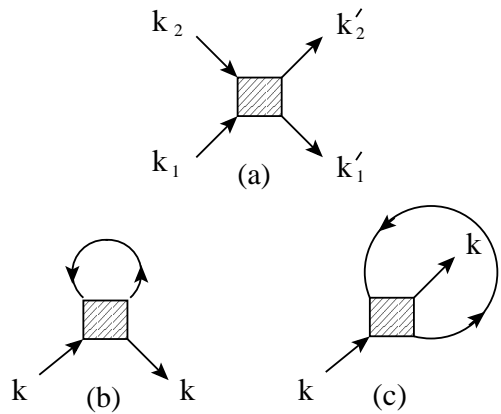}}
\caption{(Color online)  (a) Two-body composite fermion interaction with the matrix element $M_{{\bf k}_1^\prime,{\bf k}_2^\prime}^{{\bf k}_1,{\bf k}_2}$, Eq.~(\ref{matrix-element}). (b) Hartree and (c) Fock  ways of pairing the operators in the interaction Hamiltonian (\ref{second-quantized}).}
 \label{hf}
\end{figure}

One can now treat the interaction term in Eq.~(\ref{second-quantized}) in the Hartree-Fock approximation by pairing one creation and one annihilation fermionic operator $\langle  c^\dagger_{{\bf k}^\prime} c_{{\bf k}}\rangle =
\delta({\bf k}^\prime - {\bf k}) n_{{\bf k}^\prime}$.  The Hartree and Fock ways of pairing are illustrated in Fig.~\ref{hf}a,b, correspondingly.
This way one obtains the mean-field single-particle Hamiltonian
\be
                                                         \label{HF}
\hat H_\mathrm{HF} = \sum\limits_{\bf k} K_\mathrm{HF}({\bf k})\, c^\dagger_{\bf k} c_{\bf k}\,,
\ee
where the Hartree-Fock kinetic energy is given by
\be
                                                            \label{HF-energy}
K_\mathrm{HF}({\bf k})= \varepsilon_{{\bf k},-} + \frac{1}{16m} \sum\limits_{{\bf k}^\prime}
\left[ M_{{\bf k},{\bf k}^\prime}^{{\bf k},{\bf k}^\prime} -
       M_{{\bf k},{\bf k}^\prime}^{{\bf k}^\prime,{\bf k}} \right]  n_{{\bf k}^\prime},
\ee
and the self-consistency condition is imposed by requiring  $n_{{\bf k}^\prime}=f((K({{\bf k}^\prime})-\mu_F)/T)$ is the equilibrium occupation number of the state ${\bf k}^\prime$ determined by its {\em total} energy $K({{\bf k}^\prime})$ and the chemical potential $\mu_F$. The latter is to be found from particle conservation $\sum_{{\bf k}^\prime} n_{{\bf k}^\prime} =n$.

%%%%%%%%%%%%%%%%%%%%%%%%%%%%%%%%%%%%%%%%%%%%%%%%%%%%%%%%%%%%%%%
%%%%%%%%%%%%%%%%%%%%%%%%%%%%%%%%%%%%%%%%%%%%%%%%%%%%%%%%%%%%%%%%%

In the limit of small density (\ref{density-small}) we anticipate the nematic state, discussed qualitatively in section \ref{sec:fermions}. Choosing its center to be along the positive $x$-direction, one may approximate $\mathrm{arg}({\bf k})\approx k_y/k_0\ll 2\pi$. We can now employ the form-factors (\ref{form-factors}) as well as the lower branch dispersion $\varepsilon_{{\bf k},-}$ in the form of Eq.~(\ref{appspec}), to find for the Hartree-Fock energy
 \be
                                                            \label{HF-energy-1}
K_\mathrm{HF}({\bf k}) \approx  E_0 + \frac{(k_x-k_0)^2}{2m} + \frac{k_y^2}{2m_y}\,,
\ee
where
\be
                                                               \label{my}
m_y =m\,  \frac{4 k_0^2}{gn}\, \gg m
\ee
and $E_0= g\sum_{{\bf k}^\prime} (k_y^\prime)^2  n_{{\bf k}^\prime}/(4mk_0^2)$ with the effective interaction constant $g=(g_0+g_2)|c|^2 + 2(g_0-g_2)|d|^2$.
We note here that $m_y$ is essentially insensitive to the distribution $n_k$, as long as it covers a small fraction of the Rashba circle in the momentum space.
It is clear that
the Fermi surface, corresponding to the dispersion relation (\ref{HF-energy-1}) is an ellipse elongated along the $y$ direction
and centered around $(k_0,0)$, Fig.~\ref{fs}. One can now find the chemical potential of interacting fermions at $T=0$ by solving
$\sum_{K({\bf k})<\mu_F}=n$. This way we find for the chemical potential of the interacting composite fermions
\be
                                                                \label{mu-F}
\mu_F= {\frac{ 5 \pi}{ 4}}\sqrt{g}\,\, \frac{n^{3/2}}{m k_0}\,
\ee
and $E_0 = \mu_F/5$, in agreement with the estimate in the end of Section \ref{sec:int}.

\subsection{Chern-Simons gauge field}
\label{sec:CS}

So far we have been discussing the interaction energy of the composite fermions. The latter are related to bosons through the Chern-Simons
phase according to Eq.~(\ref{CB-11}). We need to discuss now the role of this phase.
At the Hartree-Fock level the system is effectively described  by non-interacting quasiparticles with the anisotropic dispersion relation Eq.~(\ref{HF-energy-1}). In the coordinate representation the momentum operators are given by $k_{x,y}\to i \partial_{x,y}$.
To bring it into a more familiar form one may shift the momentum origin to $(k_0,0)$ by the gauge factor $e^{ik_0x}$ and rescale the variables as
$\tilde x=\alpha x$ and $\tilde y= y/\alpha$, where $\alpha=(m/m_y)^{1/4}$. In the rescaled coordinates the Hartree-Fock Hamiltonian is  isotropic with the cyclotron mass $m_c=\sqrt{m m_y}$. Upon acting  on the Chern-Simons phase the rescaled momentum operator
results in a vector potential ${\bf A}(\tilde {\bf r}) $, which is included by replacing components of $i\nabla_{\tilde {\bf r}}$ by the corresponding components of $i\nabla_{\tilde {\bf r}}-{\bf A}(\tilde {\bf r})$,
\be
                                                                \label{CS}
H_\mathrm{HF} =
%\int d{\bf r} c^\dagger_{{\bf r}}
\sum\limits_{j=1}^N
{1\over 2m_c} \left[(i\partial_{\tilde x_j}-{\bf A}_x(\tilde {\bf r}_j))^2 + (i\partial_{\tilde y_j}-{\bf A}_y(\tilde {\bf r}_j))^2   \right],
\ee
where the vector potential is given by\cite{NN}
\be
                                                \label{vector-potential}
{\bf A}_\alpha(\tilde {\bf r}_j) = \sum\limits_{i\neq j}
\epsilon_{\alpha\beta}\, \frac{(\tilde {\bf r}_j-\tilde {\bf r}_i)_{\beta}}{|\tilde {\bf r}_j-\tilde {\bf r}_i|^2}\,,
\ee
while the corresponding Chern-Simons magnetic field directed perpendicular to the 2D plane is given by
\be
                                                                \label{Chern-Simons-magn-field}
B(\tilde {\bf r}_j) = \mathrm{curl}{\bf A}(\tilde {\bf r}_j) = { 2\pi}\!\sum\limits_{i\neq j} \delta(\tilde {\bf r}_j-\tilde {\bf r}_i) = { 2\pi} \!\sum\limits_{i\neq j} \delta( {\bf r}_j- {\bf r}_i)\,.
\ee
To obtain the isotropic form of the Hamiltonian acting in the {\em fermionic} space, it is important that the Chern-Simons phase was defined with the rescaled coordinates $\tilde {\bf r}$, Eq.~(\ref{CB-11}). We can now specify the value of the rescaling parameter $\alpha=(m/m_y)^{1/4}\propto (n/n_0)^{1/4}\lesssim 1$. Notice that
since $m_y$ is itself weakly dependent on $\alpha$ (through interaction form-factors $c$ an $d$) the above definition of $\alpha$ is actually a self-consistent equation. Moreover, inclusion of the magnetic field affects the wavefunctions of the quasiparticles, e.g. a homogeneous field results in the Landau quantization. That, in turn, modifies the matrix elements of interactions $M_{{\bf k}_1^\prime,{\bf k}_2^\prime}^{{\bf k}_1,{\bf k}_2}$ defining $m_y$. However, because of the form of interaction  $ F_{{\bf k}_1,{\bf k}_2}$ and $G_{{\bf k}_1,{\bf k}_2}$ the effective mass appears to depend only weakly on the specific form of the wave functions, as long as those composed of plane waves with wave vectors in the vicinity of the point $(k_0, 0)$. This appears to be the case even for a quantizing magnetic field corresponding to a fully occupied single Landau level. Therefore we use Eq.~(\ref{my}) for $m_y$ in the following and apply the effective description Eq.~(\ref{CS}) even for a quantizing magnetic field.

The importance of bringing the effective fermionic Hamiltonian to the isotropic form (\ref{CS}) is to argue that the kinetic energy does {\em not} contain large logarithms.  As explained below Eq.~(\ref{log-singular}), the kinetic energy in presence of the singular magnetic field (\ref{Chern-Simons-magn-field}) contain logarithmic integrals. At small distance they are cut by the scale of the correlation hole in the fermionic wavefunction $\Psi_F$, while at large distance they are cut by the average distance between the particles. In the isotropic fermionic state described by the Hamiltonian (\ref{CS}) both of these two length scales are given by $k_F^{-1}\sim n^{-1/2}$. As a result, the logarithms  contain a number rather than a density-dependent parameter.
Notice that without rescaling of CS phase such logarithms would contain $\log (1/\alpha)$, making the kinetic energy of the order $n^{3/2}\log (n_0/n)$. Rescaling  avoids this large extra factor in the kinetic energy.

Keeping the kinetic energy to be $\propto n^{3/2}$ by the fermionic nature of the state along with the careful rescaling of CS phase, suggests to employ
the mean-field treatment\cite{Jain,Halperin} of CS magnetic field.  It substitutes the collection of flux lines (\ref{Chern-Simons-magn-field}) by a uniform magnetic field with the same total flux. The latter is given by 2D density of particles (it is important that the rescaling of CS phase is area-preserving and thus not affecting the total flux),
\be
                                 \label{CS-mean field}
B(\tilde {\bf r}_j)\to  B ={ 2\pi} n\,.
\ee
The corresponding mean-field vector potential may be chosen as $  {\bf A}_x (\tilde {\bf r}_j) = -{\pi}n\tilde y_j$ and $ {\bf A}_y (\tilde {\bf r}_j) = {\pi}n \tilde x_j$.
In this approximation the Hamiltonian (\ref{CS}) represents a one-body problem of a particle with the cyclotron  mass $m_c=\sqrt{m m_y} \propto m \sqrt{n_0/n}$ in a constant magnetic field (\ref{CS-mean field}). The corresponding Landau spectrum is $\varepsilon_l=\omega_c(l+1/2)$, where $l=0,1,\ldots$ and the cyclotron frequency $\omega_c=2\pi n/m_c$. Since by construction we introduced exactly one flux quantum per particle, the
corresponding  filling factor is $\nu=1$.
The $N$-body groundstate  wavefunction  $\Psi_F(\tilde {\bf r}_1,\ldots,\tilde {\bf r}_N)$  is thus given by the Slater determinant built from the single-particle wavefunctions of the {\em fully occupied} lowest Landau level (LLL) (the center of mass shift in the direction of the nematic order produces an additional  multiplicative factor $\exp\{ ik_0\sum_j x_j\}$). In terms of rescaled complex coordinates $\tilde z_j=\tilde x_j+i\tilde y_j=\alpha x_j+iy_j/\alpha$ the corresponding fermionic wavefunction takes the form\cite{jbook}:
\be
                                    \label{PsiF}
\Psi_F= C_N \prod\limits_{i<j}^N (\bar{\tilde z}_i - \bar {\tilde z}_j)\,\, e^{-\pi n\sum\limits_j^N |\tilde z_j|^2/2}\, e^{ik_0\sum\limits_j^N x_j},
\ee
where $C_N$ is normalization factor. The Chern-Simons phase may be also written in terms of the rescaled complex coordinates as
$\prod (\tilde z_i-\tilde z_j)/|\tilde z_i-\tilde z_j|$. As a result the
minimal-spin component of the bosonic wavefunction (\ref{CB-11}) takes the simple form
\be
\label{CB-12}
\Psi_{\downarrow\ldots\downarrow} = 2^{-N/2}C_N \prod\limits_{i<j}^N |{\tilde z}_i -  {\tilde z}_j| \,\, e^{-\pi n\sum\limits_j^N |\tilde z_j|^2/2}\, e^{ik_0\sum\limits_j^N x_j}.
\ee
The higher spin components are obtained by acting with the spin raising operators on the minimal spin component, Eq.~(\ref{someup}). Notice that the spin-raising operator (\ref{kernel1}) is to be written in non-rescaled original coordinates ${\bf r}_j$ to ensure that the wavefunction is projected on the lower SO branch. The wavefunction (\ref{CB-12}) is independent of the CS chirality $\lambda$. It depends on the single density- and interaction-dependent parameter $\alpha=(m/m_y)^{1/4}\propto (n/n_0)^{1/4}$, which specifies the anisotropy.
This wavefunction describes an ellipsoidal droplet of the gas elongated along the $x$ direction with the ratio of $x$ and $y$ axes given by $\alpha^2$. The average density in this droplet is $n$, while its size depends on the number of particles $N$. The state (\ref{CB-12})
clearly breaks rotation symmetry. It also breaks time-reversal symmetry as well as parity. Indeed, the parity operation transforms it into the state
descendant from the maximal spin component, which is clearly a different, degenerate state. The average energy per particle for the Hamiltonian (\ref{CS}) in such variational $N$-body state is given by
\be
                                               \label{mu-result}
\mu_F = E_0+\varepsilon_0 = {\frac{3 \pi}{4}}  \sqrt{g}\,\, \frac{n^{3/2}}{m k_0}\,.
\ee
%This is the chemical potential of short-range interacting bosons with Rashba spin-orbit coupling in the mean-field approximation.
It indeed confirms the expectation that the composite fermion function of Eqs.~(\ref{someup}) and (\ref{CB-11}) yields the energy lower than TRSB and SDW bosonic states.
%%In accordance with the initial expectation it appears to be indeed smaller than the for the corresponding fermionic system, Eq.~(\ref{mu-F}).

%%%%%%%%%%%%%%%%%%%%%%%%%%%%%%%%%%%%%%%%%%%%%%%%%%%%%%%%%%%%%%%%%%%%%%%%%%%%%%%%%%%%%%%%%%%%%%%%%%%%%%%%%%%%%%%%%%
%%%%%%%%%%%%%%%%%%%%%%%%%%%%%%%%%%%%%%%%%%%%%%%%%%%%%%%%%%%%%%%%%%%%%%%%%%%%%%%%%%%%%%%%%%%%%%%%%%%%%%%%%%%%%%%%%%
%%%%%%%%%%%%%%%%%%%%%%%%%%%%%%%%%%%%%%%%%%%%%%%%%%%%%%%%%%%%%%%%
%\begin{figure}[h]
%\centerline{\includegraphics[width=55mm,angle=0,clip]{minimum-circle.eps}}
%\caption{(Color online) }
% \label{MC}
%\end{figure}
%%%%%%%%%%%%%%%%%%%%%%%%%%%%%%%%%%%%%%%%%%%%%%%%%%%%%%%%%%%%%%%%%
\section{Discussion of the results}
\label{sec:discussion}

\subsection{Phase Diagram}
\label{sec:phase-diagram}

\begin{figure}[t]
\centerline{\includegraphics[width=85mm,angle=0,clip]{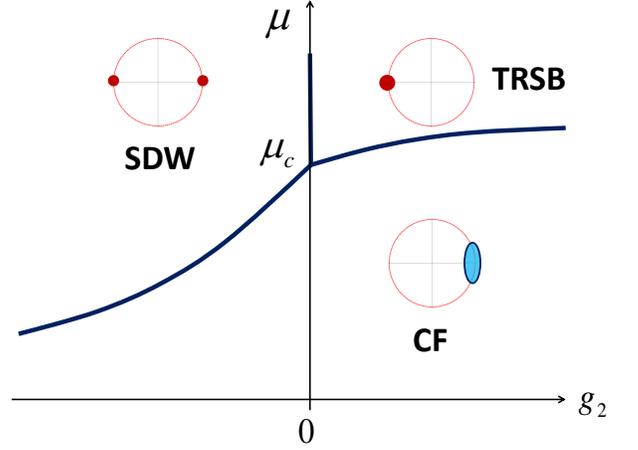}}
\caption{(Color online) Phase diagram of Rashba SO bosons.
The full lines are  first order transitions.  Insets: schematic representation of occupation factors in momentum space (red dots are coherent bosonic condensates, blue ellipse is CF Fermi sea).
 }
 \label{phase}
\end{figure}

As we have seen, the composite fermion  energy per particle is $\mu_F\sim \mu_B\sqrt{n/n_0}$, where $\mu_B$, Eq.~(\ref{mu-B}), is the chemical potential of bosonic TRSB and SDW states. Therefore at small density $n\lesssim n_0=gk_0^2$ the composite fermion (CF) groundstate  is energetically favorable over bosonic states. On the other hand, at larger density $n\gtrsim n_0$ the bosonic states, discussed earlier\cite{SAG,zhai,Wu,radic,ho11,Lamacraft,zhai2,yongping,DasSarma,ozawa,ozawa2,pitaevskii}, have lower energy. We expect thus to observe quantum phase transitions as functions of the chemical potential $\mu$ as well as anisotropy of the interactions $g_2$. The corresponding phase diagram is
schematically depicted in Fig.~\ref{phase}.

%It contains the first order transition line to SDW state at $g_2<0$ and the second order
%transition between TRSB and composite fermion (CF) states at $g_2>0$.

To find out about the nature of the transitions we recall that all three phases break rotational symmetry. In addition TRSB  breaks time-reversal,
while CF  breaks both time-reversal and parity symmetries. We expect thus the first order transition between CF and SDW, where the two discrete
symmetries $\hat T$ and $\hat P$, Eqs.~(\ref{T}), (\ref{P}), are broken simultaneously. Also the transition between SDW and TRSB, taking place close to $g_2=0$, is to be of the first order. Indeed, to avoid density wave modulation in SDW phase the populations of two opposite states on the Rashba circle must be exactly equal. Therefore the transition into TRSB phase with only single populated state must be of the first order.
%%%%%%%%%%%%%%%%%%%%%%%%%%%%%%%%%%%%%%%%%%%%%%%%%%%%%%%%%%%%%%%
%%%%%%%%%%%%%%%%%%%%%%%%%%%%%%%%%%%%%%%%%%%%%%%%%%%%%%%%%%%%%%%%%

Transition from TRSB  into CF phase breaks the parity symmetry and could be of the first or second order. One can investigate stability of the bosonic TRSB state against introduction of the small CF component.  To this end one can write a variational wavefunction, which contains
superposition of TRSB condensate, Eq.~(\ref{cond}), with $N_b$ particles and nematic CF fraction with $N_f\ll N_b$ particles and $N_b+N_f=N$,
\begin{equation}
                                                      \label{mixture}
\Psi({\bf r}_1,\ldots, {\bf r}_N\!) \propto\!\! \sum\limits_P \Psi_B^{(0)}\!({\bf r}_1,\ldots, {\bf r}_{N_b}\!)\Psi_{CF}\!({\bf r}_{N_b+1},\ldots, {\bf r}_{N}\!),
\end{equation}
where $P$ stays for a permutation of arguments between $N_b$ and $N_f$ sets and we have suppressed spin indices for brevity.
It is important to realize that the bosonic condensate and a small nematic CF fraction prefer to be on the {\em opposite} sides of the Rashba circle, Fig.~\ref{phase}. Indeed this way the spin parts of respective wavefunctions are (almost) orthogonal, minimizing the exchange energy.
The residual exchange interactions, due incomplete orthogonality, lead to CF effective mass $m_y$ in the direction tangential to the Rashba circle.
%Notice also that, since the superposition (\ref{mixture}) of the condensate and CF is incoherent, the average density is uniform irrespective of the relative occupation of the two fractions.
The energy per unit volume for the state (\ref{mixture}) is given by
\be
E= \frac{3\pi n_f^2}{4m_c} + \frac{ (n-n_f)^2}{2m}\, g_0
+ \frac{(n-n_f) n_f}{m}\, \left[g_0+g_2\right],
                                                                                                         \label{2-p}
\ee
where $n_f=N_f/N$ and the fermionic cyclotron mass $m_c=\sqrt{m m_y}=mk_0/\sqrt{(g_0-g_2)n/2}$ originates from the exchange interactions of the CF fraction with the bosonic condensate (in the limit $n_f\ll n$ the interactions between composite fermions are less important). This mean-field expression shows that for $g_2>0$ TRSB state, i.e. $n_f=0$, is always the energy minimum and thus it is stable against small CF fraction. An additional energy minimum develops at $n_f=n$  for $n<2n_0/3\pi$. Although one should not take Eq.~(\ref{2-p}) as quantitatively accurate beyond $n_f\ll n$ limit, it is true that at small enough density the CF phase costs less energy than the bosonic condensates. Together with the local stability of the condensate it suggests the {\em first} order  transition into the  CF phase.

%%%%%%%%%%%%%%%%%%%%%%%%%%%%%%%%%%%%%%%%%%%%%%%%%%%%%%%%%%%%%%%
\begin{figure}[t]
\centerline{\includegraphics[width=85mm,angle=0,clip]{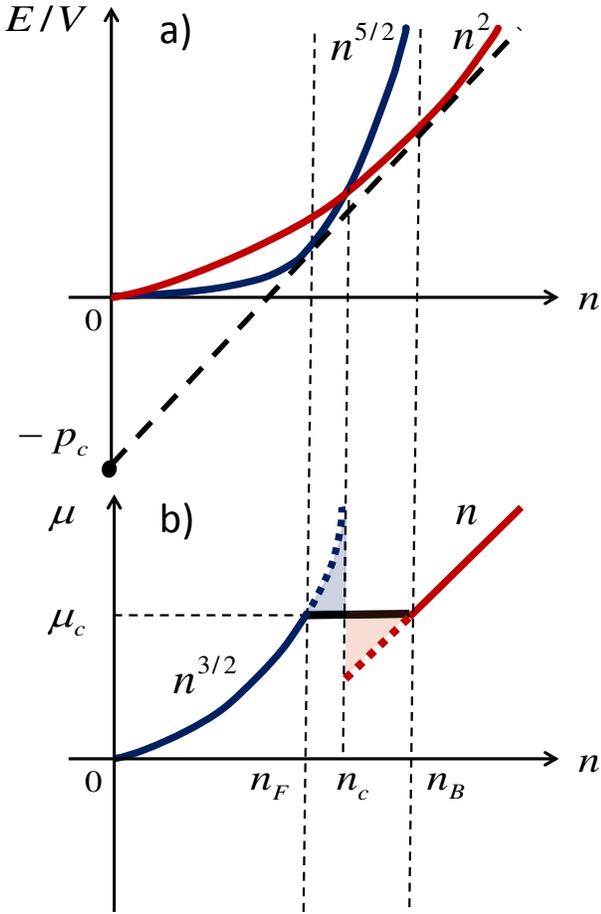}}
\caption{(Color online) (a) Specific energy vs. density for CF and Bose condensate phases. The two graphs intersects at the critical density
$n_c$. In the range $ n_F < n <n_B$ there is the phase separation. (b) Chemical potential vs. density. Maxwell construction is shown. Compare it to 1D case, Fig.~\ref{fig-1D}.}
 \label{separation}
\end{figure}
%%%%%%%%%%%%%%%%%%%%%%%%%%%%%%%%%%%%%%%%%%%%%%%%%%%%%%%%%%%%%%%%%

For the first order transition the equation of state implies the range of density where the Bose condensate (TRSW or SDW) fraction separates
from  CF fraction. This is illustrated on Fig.~\ref{separation}a. In the region $n_{F} < n < n_B$, determined by the common tangential to
$E_F(n)$ and $E_B(n)$, it is energetically favorable to spatially separate the Bose condensate with the fixed density $n_B$ from the CF liquid with fixed density $n_F$. Changing the total density within this window results in the change between the relative fraction of the two,  not changing  their individual densities. In terms of the chemical potential vs. density it means a constant critical chemical potential $\mu_c$, found from the Maxwell
construction, within the window  $n_{F} < n < n_B$, Fig.~\ref{separation}b. Alternatively, it means a range of a constant critical quantum pressure
$p_c$, Fig.~\ref{separation}a, on the pressure vs. volume $T=0$ isotherm. To minimize the interfacial energy TRSB and CF fractions
prefer to have opposite momenta and spin. This allows  avoiding paying the exchange interaction energy thanks to
orthogonality of the spinors. One expects thus to find antiferromagnetically ordered mixture of the TRSB Bose condensate and CF fractions.

\subsection{Composite fermion state in a trap}
\label{sec:trap}

Consider an axially symmetric trap created by a harmonic potential $V(r)=m\omega_0^2 r^2$. In the Thomas-Fermi approximation the local density $n(r)$ is found from the condition $V(r)+\mu(n(r))-\mu=0$, where $\mu(n)$ represents microscopic equation of state and $\mu$ is the macroscopic chemical potential, found from the condition $2\pi \int_0^R r dr n(r) =N$, where $V(R)=\mu$. For CF equation of state $\mu(n) \propto \sqrt{g}n^{3/2}/mk_0$,
this program yields:
\be
                                                   \label{trap}
\mu \propto \omega_0 N^{3/5} \left(\frac{g \omega_0 }{k_0^{2}/m}\right)^{1/5},
\quad\quad R=\sqrt{\frac{\mu}{m\omega_0^2}}\propto N^{3/10}.
\ee
Here $R$ represents the spatial extent of the $N$-particle groundstate, while $\mu$ is the typical kinetic energy per particle measured
upon trap release. These results should be compared with those for the bosonic condensate with the equation of state $\mu(n)\propto gn$. The latter yields\cite{rmp}
$\mu \propto \omega_0 (gN)^{1/2}$ and $R\propto N^{1/4}$.    The CF scaling of the chemical potential in the harmonic trap $\mu_F\propto N^{3/5}$ is valid as long as it less than the corresponding bosonic result $\mu_B\propto N^{1/2}$. Equating the two of them, one finds the condition for the number of particles within the trap below which the CF groundstate prevails
\be
                                                    \label{N-crit}
N<N_0= g^3    \left(\frac{k_0^{2}/m}{ \omega_0 }\right)^2.
\ee
This is exactly the condition of having the density in the middle of the trap less than the critical one $n(0)<n_c$.
The density profile acquires  the shape $n(r)\propto (R^2-r^2)^{2/3}$.   As opposed to the Bose condensate profile\cite{pitaevskii} $n(r) \propto (R^2-r^2)$, it exhibits infinite slop at the outer edge $r\approx R$.

In  experiments of Refs.~[\onlinecite{Lin-2008}] and [\onlinecite{Lin-2011}] synthetic gauge field and SO coupling had been engineered with the help of $\lambda = 804.1$nm Raman lasers, leading to the typical SO momentum $k_0=2\pi/\lambda$ and energy scale $k_0^{2}/m\simeq 7$kHz. Taking a trap frequency $\omega_0\simeq 30 Hz$ as in e.g. Ref.~[\onlinecite{rolston}], one obtains
$k_0^{2}/{m \omega_0 }\simeq 0.23\times 10^3$ and  $N_0\sim g^3\times 10^5$. The effective coupling constant $g$ in 2D gas of Rb with $a_s =55 \AA$ was estimated to be\cite{shlyapnikov} $g\simeq 0.2$, leading to $N_0\simeq 10^3$, which is achievable with modern detection techniques.
Interestingly enough, studying much smaller particle number experimentally
became feasible with the development of single-atom detection technology\cite{single1, single2}.

Since the CF phase has filling factor one, we expect it to be gaped in the bulk of the trap. At the surface it supports a chiral edge mode, which is similar to $\nu=1$ quantum Hall edge. In this sense CF state of Rashba SO bosons is an interacting {\em topological insulator}.

At a larger particle number
$N>N_0$ the middle of the trap has the density exceeding the critical one and thus it reverts to one of the Bose states (TRSB or SDW).
The density decays towards the edges of the trap, reaching $n_B$ at some radius. Here the phase separation, Fig.~\ref{separation},
takes place and the density discontinuously drops down to $n_F$.   The outer periphery of the trap contains ``vaporized'' low-density
CF phase, while its inner core is filled with the ``liquid'' high-density Bose condensate phase.
%The two phases are antiferromagnetically aligned.
The overall scaling of the chemical potential with the particle number approaches the Bose one, $\mu\sim N^{1/2}$.

\subsection{Rotating systems}
\label{sec:rotation}

The time-reversal and parity broken CF state is chiral. It is not immediately obvious from the minimal spin component wavefunction (\ref{CB-12}), since it depends on the absolute values $|\tilde z_i -\tilde z_j|$ only. However, the spin-raising kernel (\ref{kernel1}) is certainly chiral and so are all the higher spin components of the many-body bosonic wavefunction (\ref{someup}). The degenerate state descending from the maximal spin component (which has exactly the same  form as Eq.~(\ref{CB-12})) has the opposite chirality. The CF groundstate spontaneously breaks the $Z_2$ symmetry between them.  Rotation of the system serves as an explicit symmetry breaking perturbation, enforcing one of the sates vs. another. Indeed, rotation with the angular frequency $\Omega$ may be viewed as an external magnetic field\cite{wilkin} ${\bf B}_\mathrm{rot} = 2m_c\Omega\hat {\bf z}$, which adds up to CS magnetic field ${\bf B}=\pm 2\pi n \hat {\bf z}$. As a result the total  flux is bigger or smaller than one flux quantum per particle, depending on whether CS magnetic filed is parallel or antiparallel with the rotation direction. This either creates $m_c\Omega/\pi$ holes per unit area in  LLL, or promotes the same number of particles to the next Landau level. At least from the standpoint of the kinetic energy the former alternative requires twice less energy than the latter. One expects thus that the symmetry is broken in a way to add constructively CS and rotational fields to create holes in LLL. Due to presence of these holes, one expects gapless bulk excitations  in the rotating system.
(Alternatively in analogy with FQHE, excitations may have gaps  substantially reduced compared to the $\Omega=0$ case.)

\subsection{Fractional Hall states?}

We have employed Chern-Simons phase with one flux quantum per particle to convert composite fermions into bosons, i.e. $\lambda=\pm 1$ in Eq.~(\ref{CB-11}). One could also attach three flux quanta by choosing $\lambda=\pm 3$. This choice leads to the effective magnetic field
$B=6\pi n$ and thus to $\nu =1/3$ filling factor. The CF variational groundstate is then given by the Laughlin\cite{laughlin1} state, i.e.
$ (\bar{\tilde z}_i - \bar {\tilde z}_j) \to (\bar{\tilde z}_i - \bar {\tilde z}_j)^3$ in Eq.~(\ref{PsiF}). Upon multiplication on CS phase
with $\lambda =3$ it leads to the following expression for the minimal spin component
\be
\label{Laughlin}
\Psi_{\downarrow\ldots\downarrow} \propto \prod\limits_{i<j}^N |{\tilde z}_i -  {\tilde z}_j|^3 \,\, e^{-3\pi n\sum\limits_j^N |\tilde z_j|^2/2}\, e^{ik_0\sum\limits_j^N x_j},
\ee
notice the factor of $3\pi n$ in the exponent which ensures the correct total density $n$.
The kinetic energy of such state is factor of three higher than that ascending from Eq.~(\ref{CB-12}). On the other hand, the correlation holes
are wider, which may lead to a gain in the interaction energy (especially since the interactions due to higher spin components are essentially non-local). While at the moment we are not aware of a model where the fractional state is advantageous, it is worth pointing out that such a model is in principle possible.

Notice that symmetric minimal spin component may be written with an arbitrary integer exponent $p$ as $\propto \prod_{i<j}^N |{\tilde z}_i -  {\tilde z}_j|^p e^{-p\pi n\sum\limits_j^N |\tilde z_j|^2/2}\, e^{ik_0\sum\limits_j^N x_j}$. However only {\em odd} $p$'s may be traced to CF construction. This seems to indicate that states with even $p$'s yield larger
average energy. This is indeed the case for $p=0$, which is nothing but TRSB condensate with $\mu_B\propto n$. It is not clear at the moment how to demonstrate this statement for $p=2,4,\ldots$.

\subsection{Conclusions}

We have shown that the low-density phase of Rashba SO bosons may be described as the composite fermion state in the quantizing magnetic filed. Such a state is very different from both TRSB  Bose condensate and SDW state discussed before. In particular its equation of state $\mu(n)\propto n^{3/2}$ leads to a different scaling of the kinetic energy and atomic cloud size with the number of particles in shallow traps. It also implies different profile of the cloud density. The excitation spectrum is predicted to be gaped in the bulk of trap with the gapless chiral surface mode at its edge. For deeper traps we predict the phase separation between denser condensate phase in the middle of the trap and dilute CF phase at its edge with the first order density jump at the interface between them.

\section{Acknowledgments}

We are grateful to E. Altman, N. Cooper, G. Juzeliunas, A. Lamacraft and A. Vishwanath for useful discussions.  LG thanks Aspen Center for Physics for hospitality. This work was supported by DOE contract DE-FG02-08ER46482.

\appendix

\section{}\label{app:A}

Here we give details of  calculations
for the components of the two-particle wavefunction,
$\Psi_{ \uparrow \uparrow}({\bf r},{\bf r})$ and $\Psi_{ \uparrow \downarrow}({\bf r},{\bf r})=\Psi_{\downarrow \uparrow }({\bf r},{\bf r})$
in Eq.~(\ref{2-inter}) and
derive Eqs.~(\ref{coinciding-points}), (\ref{form-factors}) of the main text.

\subsection{Calculation of $\Psi_{ \uparrow \uparrow}({\bf r},{\bf r})$ and the interaction form-factor $F_{\bf k_1,k_2}$}

From Eq.~(\ref{2-particle2}) we have
\bea
\label{a1}
&&\Psi_{\uparrow\uparrow}({\bf r},{\bf r})=\\
&&\frac{1}{2}\int d{\bf r}_1^\prime d{\bf r}_2^\prime {\cal R}({\bf r} - {\bf r}_1^\prime) {\cal R}({\bf r} - {\bf r}_2^\prime)e^{i \lambda  \arg({\tilde{\bf r}^\prime_1}-{\tilde{\bf r}^\prime_2})} \Psi_{F}({\bf r}_1^\prime,{\bf r}_2^\prime),\nn
\eea
where the fermionic part of the wave function is given by
\bea
\label{a3}
\Psi_{F}({\bf r}_1,{\bf r}_2)=\frac{1}{\sqrt{2}V}\Big(e^{i {\bf k_1}{\bf r_1}+i {\bf k_2}{\bf r_2} }
-e^{i {\bf k_2}{\bf r_1}+i {\bf k_2}{\bf r_2} }\Big)\nn\\
\eea
and anisotropic vectors ${\tilde{ \bf r}^\prime_1}$ and ${\tilde{\bf r}^\prime_2}$ in the Chern-Simons factor $e^{i \lambda  \arg({ \tilde{ \bf r}^\prime_1}-{\tilde{\bf r}^\prime_2})} $ are defined as
${\tilde{\bf r}^\prime_j}=(\alpha x_j^\prime, y_j^\prime/\alpha)$, $j=1,2$, where $\alpha$ is the anisotropy parameter.

To represent the integrand in Eq.~(\ref{a1}) in a convenient form, we convert it to the momentum space.
Fourier transformation yields
\bea
\label{a2}
e^{i \lambda \arg {\bf \tilde{r}}} = -2\pi i \int\frac{d {\bf k}}{(2 \pi)^2}\, \frac{e^{i \lambda \arg{\tilde{\bf k}}}}{\tilde{\bf k}^2}\,
e^{i {\bf k}{\bf r}},
\eea
with ${\tilde{ \bf k}}=(k_x/\alpha ,\; \alpha k_y)$. Fourier image of ${\cal R}({\bf r})$ is ${\cal R}({\bf k})=ie^{-i \arg{\bf k}}$. Substituting now Eqs.~(\ref{a3}) and (\ref{a2}) into (\ref{a1}), we obtain
\bea
\label{a4}
&&\Psi_{\uparrow\uparrow}({\bf r},{\bf r})=-\frac{i e^{i({\bf k}_1+{\bf k}_2){\bf r}}}{4 V \sqrt{2}\pi}
\int d{\bf k}\, \frac{e^{i \lambda \arg{\tilde {\bf k}}}}{\tilde{\bf k}^2}     \\
&&\times \Big(e^{-i \arg({\bf k}_1+{\bf k})-i \arg({\bf k}_2-{\bf k})}-e^{-i \arg({\bf k}_2+{\bf k})-i \arg({\bf k}_1-{\bf k})}\Big)\nn
\eea
In order to evaluate  $F_{{\bf k}_1{\bf k}_2}$ from (\ref{a4}), it is convenient to introduce
the following notations
\bea
\label{a5}
e^{i \lambda \arg{\bf \tilde{k}}}&=&\frac{k_x/\alpha+i \lambda \alpha k_y}{|\tilde{k}|}=\left(\frac{\tilde{z}}{|\tilde{z}|}\right)^{\lambda},\nn\\
e^{-i \arg({\bf k}_1+{\bf k})}&=&\frac{\bar w_1+\bar z}{|w_1+z|},\;\; w_1=k_{1,x}+i k_{1,y},\nn\\
e^{-i \arg({\bf k}_2-{\bf k})}&=&\frac{\bar w_2-\bar z}{|w_2-z|},\;\; w_2=k_{2,x}+i k_{2,y},\nn\\
z&=& k_x+ i k_y.
\eea
The complex variable $\tilde{z}= k_x/\alpha+i \alpha k_y$  can be conveniently recast as $\tilde{z} =\frac{z}{2}(\alpha+1/\alpha)+\frac{\bar z}{2}(1/\alpha-\alpha)$. Employing  integration variables $z$ and $\bar{z}$ and
using complex representation of vectors ${\bf k}_1$ and ${\bf k}_2$, one obtains
\bea
\label{a6}
&&F_{{\bf k}_1{\bf k}_2}=\frac{i}{\sqrt{2}}\int \frac{dz d\bar z}{4\pi}\frac{1}{|\tilde z|^2}\Biggl(\frac{\tilde z }{|\tilde z |}\Biggr)^{\lambda}
\sqrt{\frac{\bar w_1 \bar w_2}{w_1  w_2}}\nn\\
&\times&\left[\frac{\frac{\bar z^2}{\bar w_1\bar w_2}-1+
\frac{\bar z(\bar w_1-\bar w_2)}{\bar w_1\bar w_2}}{\sqrt{|\frac{\bar{z}^2}{\bar w_1 \bar w_2}-1+
\frac{\bar z (\bar w_1-\bar w_2)}{\bar w_1\bar w_2}}|^2}-(w_1\longleftrightarrow w_2)\right].
\eea
In Eq.~(\ref{a6}) the short hand notation $(w_1\longleftrightarrow w_2)$ stands for the same first expression
in square brackets but with interchanged variables $w_1$ and $w_2$.

\begin{figure}[t]
\centerline{\includegraphics[width=85mm,angle=0,clip]{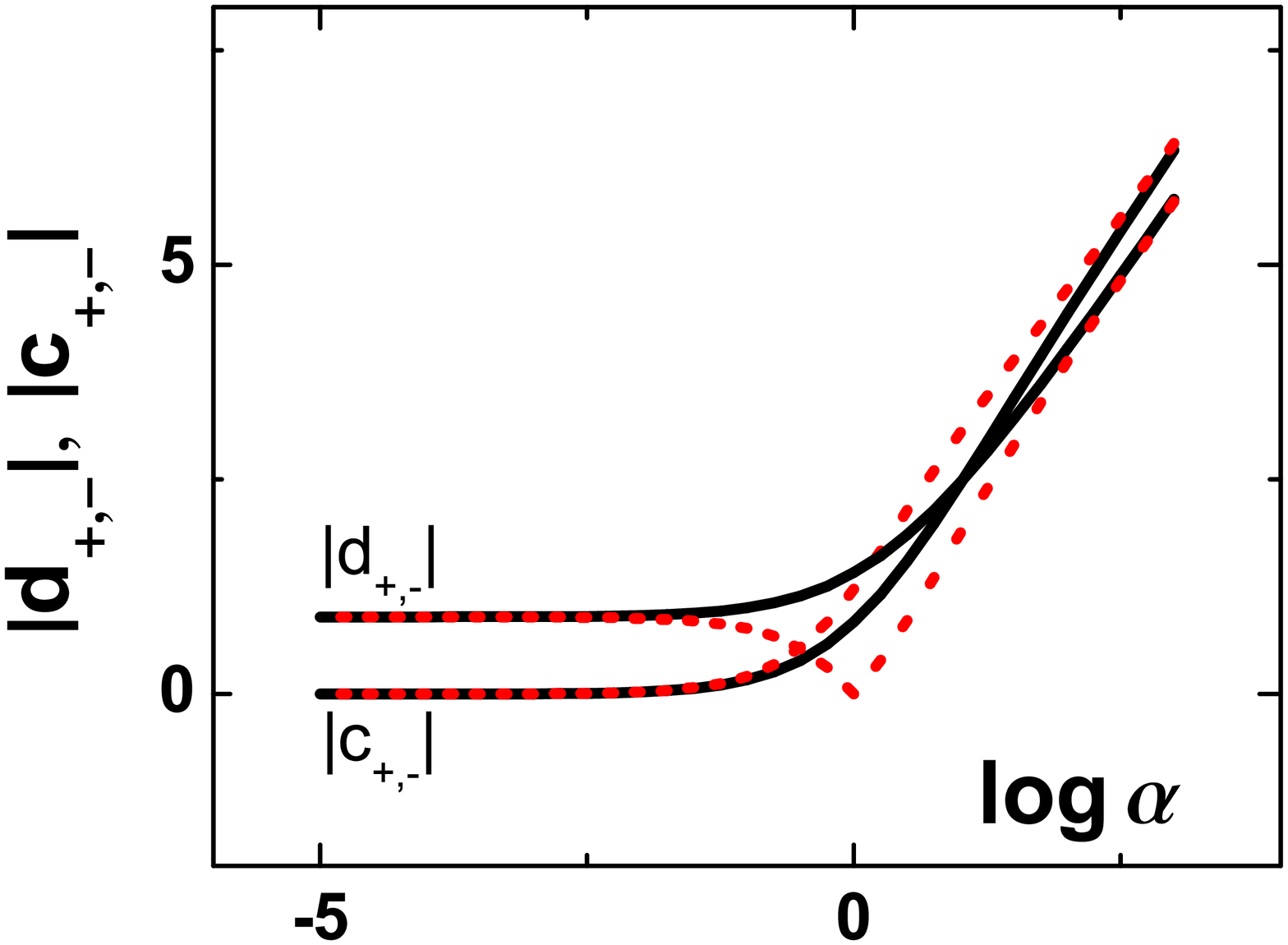}}
\caption{(Color online) Interaction coefficients $|c_\lambda(\alpha)|$   and  $|d_\lambda(\alpha)|$. Full lines correspond to $\lambda=-1$ and dotted lines -- to $\lambda=+1$ chirality.}
 \label{combined}
\end{figure}

Upon introducing new dimensionless variables
\bea
\label{a7}
\nu = \frac{z}{\sqrt{w_1 w_2}},\;\; \tau=\frac{w_1-w_2}{\sqrt{w_1 w_2}},
\eea
Eq.~(\ref{a6}) yields
\bea
\label{a8}
&& F_{{\bf k}_1{\bf k}_2}=-\frac{i}{\sqrt{2}}\sqrt{\frac{\bar w_1 \bar w_2}{|w_1 w_2|}}\int\frac{d\nu d\bar\nu}{4 \pi}\\
&&\times \Biggl(\frac{\tilde \nu}{|\tilde \nu|}\Biggr)^{\lambda}\frac{1}{|\tilde \nu|^2}\Bigg(\frac{\bar \nu^2-1+\bar \nu \bar \tau}{\sqrt{|\bar{\nu}^2-1+\bar\nu \bar \tau}|^2}-(\tau\rightarrow -\tau)\Bigg),\nn
\eea
where $\tilde{\nu}=\frac{\nu}{2}(\alpha+1/\alpha)+\frac{\bar \nu}{2}(1/\alpha-\alpha)$ and
parameter $\tau$ can be rewritten in terms of original momenta ${\bf k}_1$ and ${\bf k}_2$ as follows
\bea
\label{a8-2}
\tau&=&\sqrt{\frac{k_1}{k_2}}e^{i(\arg({\bf k}_1)-\arg({\bf k}_2))/2}
-\sqrt{\frac{k_2}{k_1}}e^{-i(\arg({\bf k}_1)-\arg({\bf k}_2))/2}.\nn\\
\eea
The integral on the right hand side of (\ref{a8}) is convergent, therefore one can expand the integrand  in $\bar \tau$ and then perform
the integration.
Keeping only linear in $\bar \tau$ terms in this expansion one finds
\be
\label{a8-21}
F_{\bf k_1,k_2}=ic\sqrt{\frac{\bar w_1 \bar w_2}{|w_1 w_2|}}\, \bar \tau\simeq  ic\left( e^{-i\arg({\bf k}_1)}-e^{-i \arg({\bf k}_2)}\right) ,
\ee
where
\be
\label{cc}
c_{\lambda}(\alpha)=\int\frac{d\nu d\bar\nu}{\sqrt{2} 4 \pi}
\Biggl(\frac{\tilde \nu}{|\tilde \nu|}\Biggr)^{\lambda}\frac{1}{|\tilde \nu|^2}
\frac{(|\nu|^2-1)({\bar \nu}^2-1)(\nu+{\bar \nu})}{|{\bar \nu}^2-1|^3},
\ee
%For $\alpha>1$ we find the function $|c_\lambda(\alpha)|^2$ in Eq.~~(\ref{form-factors}) using the symmetry property
%$|c_\lambda(\alpha)|^2=|c_\lambda(1/\alpha)|^2$.
The functions $|c_\pm(\alpha)|$ are plotted in Fig.~\ref{combined}.

%%%%%%%%%%%%%Here $\bar{g}=g_0-g_2$%%%%%%%%%%%%%%%%%%%%%%%%%%%%%%%%%%%%%%%%%%%%%%%%%%
%%%%%%%%%%%%%%%%%%%%%%%%%%%%%%%%%%%%%%%%%%%%%%%%%%%%%%%%%%%%%%%%%

\subsection{Calculation of $\Psi_{\uparrow\downarrow}({\bf r},{\bf r})$ and the interaction form-factor $G_{\bf k_1k_2}$}

From Eq.~(\ref{2-particle2}) one finds
\bea
\label{a10}
\Psi_{\uparrow\downarrow}({\bf r},{\bf r})&=&\frac{1}{2}\int d{{\bf r}_1^\prime}  {\cal R}({\bf r} - {\bf r}_1^\prime)
e^{i\lambda \arg({\bf \tilde{r}}^\prime_1-{\bf \tilde{r}})} \Psi_{F}({\bf r}_1^\prime,{\bf r}),\nn\\
\eea
where $\Psi_{F}({\bf r}_1^\prime,{\bf r})$ is defined in Eq.~(\ref{a3}), while spin-rising operator  ${\cal R}$
is given by Eq.~(\ref{kernel1}). After substituting these expressions into (\ref{a10}) we obtain
\bea
\label{a11}
&&\Psi_{\uparrow\downarrow}({\bf r},{\bf r})=  \frac{e^{i({\bf k}_1+{\bf k}_2){\bf r}}}{2 V }G_{\bf k_1k_2},\\
&&G_{\bf k_1k_2}=-\frac{1}{\sqrt{2}\pi}\!\int\!
 d{\bf r}^\prime \frac{e^{i\lambda \arg({\bf \tilde r}-{\bf \tilde r'})-i \arg({\bf r}-{\bf r'})} }{|{\bf r} - {\bf r}^\prime|^2}\nn\\
&&\times\Big(\! e^{-i {\bf k}_2({\bf r}-{\bf r}^\prime) }
\! - \! e^{-i {\bf k}_1({\bf r}-{\bf r}^\prime) }\!\Big) \nn
\eea
which shows that $G_{\bf k_1k_2}$ is a difference of two functions $G_{\bf k_1k_2}=\tilde{I}_{\bf k_1}-\tilde{I}_{\bf k_2}$.
Introducing $({\bf r-r'})_x+i ({\bf r-r'})_y= r e^{i \phi}$, and $\beta=\log(\alpha)$, we will have  for $\lambda=+1$ chirality
\bea
\label{a11-1}
e^{i\arg({\bf \tilde r}-{\bf \tilde r'})}&=&\frac{\cosh[\beta] e^{i \phi}+\sinh[\beta] e^{-i \phi}}{|\cosh[\beta]
e^{i \phi}+\sinh[\beta] e^{-i \phi}|}\\
&=&\frac{\cosh[\beta] e^{i \phi}+\sinh[\beta] e^{-i \phi}}{\sqrt{ \cosh[2 \beta](1+\tanh[2\beta] \cos[2 \phi])}}.\nn
\eea
Importantly, the form-factor corresponding to the  $\lambda=-1$ chirality can be obtained
by interchanging $\cosh[\beta]$ with $\sinh[\beta]$ in the numerator of Eq.~(\ref{a11-1}).
Then we obtain
\bea
\label{Ik}
&&\tilde{I}_{\bf k}=\frac{1}{\pi \sqrt{2 \cosh[2\beta]}}\\
&\times &\int\frac{d r}{r}d\phi \frac{(\cosh[\beta]+\sinh[\beta] e^{-2 i \phi})e^{-i k r\cos[\phi-\arg({\bf k})]}}
{\sqrt{1+\tanh[2\beta] \cos[2 \phi]}}.\nn
\eea
We note that the integral $\tilde{I}_{\bf k}$ has a logarithmical divergent contribution coming from  $r\rightarrow 0$, but it cancels out in the difference  $\tilde{I}_{\bf k_1}-\tilde{I}_{\bf k_2}$
and therefore in $G_{\bf k_1k_2}$.
In order to  find the coefficient $d_{\lambda}(\alpha)$ in Eq.~(\ref{form-factors}) we need
to evaluate  integral (\ref{Ik}) in linear over $\arg({\bf k})\ll 2\pi $ approximation. Upon expanding integrand over $\arg({\bf k})$ and
integrating over $r$  we arrive to the following expression
\bea
\label{dd1}
d_{\lambda}(\alpha)=\frac{i s_{\lambda}}{\pi \sqrt{2 \cosh[2\beta]}}
\int_0^{2 \pi} d\phi \frac{1-\cos[2 \phi]}{\sqrt{1+\tanh[2\beta] \cos[2 \phi]}},\nn\\
\eea
where  $s_{+}=\sinh[\beta]$  and  $s_{-}=\cosh[\beta]$.
The function $|d_\pm(\alpha)|$ is plotted vs $\ln\alpha$ for both chiralities $\lambda=\pm 1$
 in Fig.~\ref{combined}.

\end{document}